\documentclass[a4paper,11pt]{article}
\pdfoutput=1 

\usepackage{jheppub} 

\usepackage[T1]{fontenc} 
\usepackage{physics}
\usepackage{slashed}
\usepackage{natbib}
\usepackage{mathtools}

\title{\boldmath Fermion-Number Variation for Several Coupling Schemes in an $\mathrm{SU}\qty(N)$ Chiral Soliton Background}

\author[a]{Nobuyuki Sawado,}
\author[a,1]{Shintaro Yamamoto\note{Corresponding author.}}

\affiliation[a]{Department of Physics and Astronomy, Faculty of Science and Technology, Tokyo University of Science, Noda, Chiba 278-8510, Japan}

\emailAdd{sawadoph@rs.tus.ac.jp}
\emailAdd{shintaroyamamoto019@gmail.com}

\abstract{
When topological solitons are coupled to fermions, they often induce a nontrivial fermion number. 
In the chiral quark soliton model (CQSM), this phenomenon underlies the identification of soliton topology with baryon number. 
However, whether this topology--fermion-number correspondence is determined solely by soliton topology or also depends on the fermion--soliton coupling scheme remains unclear. 
In this paper, we investigate this question by comparing several coupling schemes to the same $\mathrm{SU}\qty(N)$ chiral soliton background. 
To treat the corresponding Dirac Hamiltonians within a common framework, we evaluate the variation of the regularized fermion-number along a model-parameter space. 
For the standard chiral Yukawa coupling of the CQSM, this variation reproduces the topological charge $Q$ associated with $\pi_3\qty(\mathrm{SU}\qty(N))$. 
The same result is obtained for an axial-vector coupling, whereas the fermion number variation vanishes for a vector coupling, despite the nontrivial topology of the chiral soliton background. 
These results provide an explicit example in which soliton topology alone does not determine the fermionic response; the topology--fermion-number correspondence also depends on how the background field is coupled to fermions.
}

\begin{document} 
\maketitle
\flushbottom

\section{Introduction}
\label{Introduction}

Quantum chromodynamics (QCD) describes the fundamental interaction between quarks and gluons. 
At low energies, however, the strong-coupling nature of QCD makes its direct treatment highly challenging. 
To address this, the chiral quark soliton model (CQSM) was proposed in the 1980s as a low-energy effective theory of QCD~\cite{Alkofer:1995mv,Diakonov:1987ty,Diakonov:1997sj,Diakonov:2000pa}. 
Since then, the CQSM has been extensively studied as an effective description of baryons and successfully applied to various phenomenological problems, including the spectroscopy of octet and decuplet baryons, parton distribution functions, exotic baryons such as pentaquark candidates, and more recently, heavy baryons containing charm and bottom quarks~\cite{CHRISTOV199691,Diakonov:1997sj,Yang:2004,Wakamatsu:1990ud,Wakamatsu:1998rx,Diakonov:1997mm,Suh:2022}. 
Among the characteristic features of the CQSM, we focus on the relationship between the topology of the chiral soliton and the fermionic spectrum. 

In the CQSM, the quark field is coupled to a chiral field through a chiral Yukawa coupling. 
The Lagrangian in $\qty(3+1)$-dimensional Minkowski spacetime is given by
\begin{align}
	\mathcal{L}=\overline{\Psi}\qty(i\gamma^\mu\partial_\mu-MU^{\gamma_5})\Psi,\ U^{\gamma_5}\equiv \frac{1+\gamma_5}{2}U+\frac{1-\gamma_5}{2}U^\dagger\label{CQSM}
\end{align}
where $U\in\mathrm{SU}\qty(N)$ is a chiral field coupled to the quarks, and $M$ is the dynamical quark mass arising from spontaneous chiral symmetry breaking. 
The one-loop effective action of Eq.~\eqref{CQSM} admits a soliton solution, known as the skyrmion~\cite{Manton:2004tk,Shnir:2018yzp,Manton:2022fcb,Diakonov:1997sj}. 
Solving the Dirac equation in the skyrmion background yields the corresponding fermionic spectrum. 
As the dynamical quark mass $M$ is varied, the energy levels in this spectrum exhibit characteristic level-crossing behavior, known as spectral flow. 
In particular, the valence level crosses the zero-energy line, and eventually dives into the negative-energy continuum (Dirac-sea)~\cite{Kahana:1984be,Kahana:1984dx,Sawado:1998gk,Sawado:2002zm,Sawado:2004pm}. 
This spectral flow is closely related to the topology of the chiral soliton. 
Specifically, the chiral field $U$ defines a map from compactified physical space $S^3$ to the target space $\mathrm{SU}(N)$, and the corresponding homotopy group $\pi_3(\mathrm{SU}(N))\simeq\mathbb{Z}$ characterizes this map by an integer, the topological charge $Q\in\mathbb{Z}$. 
In the CQSM, this topological charge is reflected in the net number of energy levels crossing the zero-energy line in the fermionic spectrum~\cite{Diakonov:1997sj,Niemi:1984di,Niemi:1984vz,Kahana:1984be,Kahana:1984dx,Sawado:1998gk,Sawado:2002zm,Sawado:2004pm}. 
Similar phenomena have been extensively investigated in various fermion--soliton systems~\cite{Jackiw:1976,Jackiw:1981ee,Callan:1982ac,Rubakov:1988aq,Gani:2022ity,Perapechka:2018yux,Perapechka:2019upv,Amari:2024rpm,Amari:2026,Dzhunushaliev:2024kti,Kahana:1984be,Kahana:1984dx,Sawado:1998gk,Sawado:2002zm,Sawado:2004pm}. 
In this broader context, the correspondence between the topology of solitons and fermionic spectrum is closely related to the Cheshire Cat picture~\cite{ALKOFER1996139,Alkofer:1995mv,Zahed:1986qz}, which provides an interpretation of baryon number as a topological quantity associated with the chiral field~\cite{Diakonov:1997sj,Witten:1983tw,Witten:1983tx}. 
In other words, the spectral flow provides the underlying mechanism by which the topological charge of solitons is transferred to the fermionic spectrum, yielding a nontrivial fermion number~\cite{Diakonov:1997sj,Niemi:1984di,Niemi:1984vz}. 

Despite the widespread occurrence of such a topology--fermion-number correspondence, whether the soliton topology alone determines the correspondence remains unclear. 
Within the CQSM, the chiral Yukawa coupling in Eq.~\eqref{CQSM} is the standard coupling between the Dirac and chiral fields. 
However, the same chiral field $U$, with the same topological charge $Q$, can couple to fermions in several ways, e.g., through a vector coupling or an axial-vector coupling. 
Since these couplings define different fermionic Hamiltonians, whether they yield the same induced fermion number for the chiral soliton remains unclear a priori. 
This ambiguity motivates the present study's central question: {\it whether the topology--fermion-number correspondence depends on the fermion--soliton coupling scheme}. 
Understanding such a dependence provides a first step toward classifying which coupling schemes allow the topological fermion number. 
In the CQSM, this viewpoint may also clarify the topological significance of the chiral Yukawa coupling associated with spontaneous chiral symmetry breaking. 

Accordingly, to address this question, we compare the induced fermion number for several coupling schemes with a common chiral field. 
For general soliton backgrounds, this quantity has typically been evaluated using derivative expansions~\cite{Goldstone:1981,Diakonov:1997sj,Niemi:1984vz,Diakonov:1987ty,Niemi:1984dp}. 
In the present work, however, we employ another approach that facilitates a comparison of the fermion-number responses. 
Specifically, we define the regularized fermion number and evaluate its variation with the dynamical quark mass $M$, considering $M=M_0$ as a reference point at which the fermion number vanishes. 
The resulting variation therefore quantifies the fermion number relative to this reference state. 
This formulation, in turn, provides a convenient framework for comparing the fermion-number responses among different coupling schemes.

As a benchmark, we first apply this method to the standard chiral Yukawa coupling in the CQSM, because the associated topology--fermion-number correspondence is well established. 
For this coupling, the fermion-number variation reproduces the topological charge $Q$ of the chiral field. 
We then consider an axial-vector coupling and a vector coupling constructed from the same chiral field $U$. 
For the axial-vector coupling, the fermion-number variation likewise reproduces the topological charge $Q$, whereas for the vector coupling, the variation vanishes despite the nontrivial topological charge of $U$. 
Within the formal expansion employed in our analysis, we further show, term by term, that the remaining terms do not contribute to the fermion-number variation. 
Overall, these results provide an explicit example indicating that the topology of the soliton background alone does not determine the fermion number. 
Instead, the realization of the topology--fermion-number correspondence depends on the fermion--soliton coupling scheme. 

This paper is organized as follows. 
In Sec.~\ref{Model}, we briefly introduce the chiral soliton background of the CQSM and the associated topological charge. 
In Sec.~\ref{General}, we formulate the regularized fermion number and define its variation with the dynamical quark mass $M$. 
In Sec.~\ref{Comp}, we evaluate the fermion-number variation for the standard chiral Yukawa coupling and establish its relation to the topological charge of the chiral soliton background. 
In Sec.~\ref{Comp2}, we extend the analysis to the axial-vector and vector couplings and compare their responses. 
Finally, Sec.~\ref{Summary} is devoted to the summary and discussion.


\section{Chiral Soliton Background and Its Topological Charge}
\label{Model}

In this section, we briefly introduce the chiral soliton background and the corresponding Dirac Hamiltonian in the CQSM. 
We begin with the partition function 
\begin{align}
	\mathcal{Z}
	&=\int\mathcal{D}U\mathcal{D}\Psi\mathcal{D}\overline{\Psi}\exp\qty[i\int\dd[4]{x}\overline{\Psi}\qty(i\gamma^\mu\partial_\mu-MU^{\gamma_5})\Psi].\label{partition}
\end{align}
Here $\Psi$ denotes an $\mathrm{SU}\qty(N)$-flavor multiplet of Dirac fields and $\overline{\Psi}$ is defined by $\overline{\Psi}\equiv\Psi^\dagger\gamma^0$. 
The gamma matrices satisfy the Clifford algebra $\qty{\gamma^\mu,\gamma^\nu}=2\eta^{\mu\nu}$ with the Minkowski metric $\eta^{\mu\nu}=\mathrm{diag}\qty(1,-1,-1,-1)$ and also the relations
\begin{align}
	\qty(\gamma^0)^\dagger=\gamma^0,\ \qty(\gamma^j)^\dagger=-\gamma^j,\ \qty(\gamma^5)^\dagger=\gamma^5=\gamma_5=i\gamma^0\gamma^1\gamma^2\gamma^3.
\end{align}
For later use, we define the chirality projection operators $P_\pm\equiv\frac{1\pm\gamma_5}{2}$ and express the chiral matrix $U^{\gamma_5}$ as $U^{\gamma_5}\equiv UP_++U^\dagger P_-$, where $U\in\mathrm{SU}\qty(N)$ is the chiral field. 
The dynamical quark mass $M$ is a model parameter and quantifies the degree of chiral symmetry breaking. 
Since the CQSM provides an effective description of the spontaneously broken phase, we restrict ourselves to the broken-symmetry region $M>0$. 
Within this region, a natural one-parameter family of Dirac operators is provided by gradually increasing $M$ and turning on the coupling strength. 
In the following, we consider the parameter space 
\begin{align}
	M\in\left[M_0,\infty\right),\ M_0>0,\label{parameterspace}
\end{align}
where $M_0>0$ denotes a reference point\footnote{For the standard self-consistent $\mathrm{SU}\qty(2)$ CQSM, the critical dynamical quark mass for the existence of solitonic solutions is typically $350\ \mathrm{MeV}$~\cite{Diakonov:1997sj,CHRISTOV199691}. Here $M_0$ should be understood only as a reference scale of this order.}.

By integrating out the Dirac field in Eq.~\eqref{partition}, one obtains the effective action 
\begin{align}
	S_{\mathrm{eff}}=-n_c\ln\det i\slashed{\mathfrak{D}}
\end{align}
where $i\slashed{\mathfrak{D}}=i\gamma_\mu\partial_\mu+MU^{\gamma_5}$ denotes the Dirac operator in Euclidean spacetime and $n_c$ is the color number (see also~\cite{Diakonov:1997sj}), i.e., the color degrees of freedom yield an overall $n_c$-fold degeneracy. 
Since $i\slashed{\mathfrak{D}}$ is not a hermitian operator, the effective action contains the imaginary part: $S_{\mathrm{eff}}=\Re S_{\mathrm{eff}}+i\Im S_{\mathrm{eff}}$. 
By expanding the real part in terms of $\partial_\mu U$, and performing the derivative expansion~\cite{Diakonov:1987ty,Diakonov:1997sj}, up to quartic terms, we obtain 
\begin{align}
	\Re S_{\mathrm{eff}}
	&=\frac{n_cF_\pi^2}{4}\int\dd[4]{x}\mathrm{Tr}\qty(\partial_\mu U^\dagger\partial_\mu U)\nonumber\\
	&+\frac{n_c}{192\pi^2}\int\dd[4]{x}\mathrm{Tr}\qty{-\qty[U^\dagger\partial_\mu U,U^\dagger\partial_\nu U]^2+\qty(U^\dagger\partial_\mu UU^\dagger\partial_\nu U)^2}\label{ReS}
\end{align}
where $F_\pi$ is the pion decay constant. 
The first two terms of Eq.~\eqref{ReS} have the same structure\footnote{It is known that the remaining fourth derivative term may destabilize the classical soliton solution~\cite{Dhar:1985,AITCHISON1985162}, which can be understood using Derrick's scaling argument~\cite{Derrick:1964}. However, based on the similarity between the effective action and the Skyrme model, it has been recognized that the chiral field in the CQSM has a skyrmion-type configuration.} as the Skyrme model~\cite{Manton:2004tk,Manton:2022fcb}. 
In this sense, the finite energy static configuration of the effective action is called a skyrmion. 
For finiteness of the energy, we impose the boundary condition $U\qty(x)\to I_N$ as $\abs{x}\to\infty$, where $I_N$ is the $N\times N$ identity matrix. 
This implies a one-point compactification of physical space $\mathbb{R}^3$, so that topologically $\mathbb{R}^3\cup\qty{\infty}\simeq S^3$, and then $U$ is a map from $S^3$ to $\mathrm{SU}\qty(N)$. 
As is well known, such maps are classified by the homotopy group $\pi_3\qty(\mathrm{SU}\qty(N))\simeq\mathbb{Z}$~\cite{Manton:2004tk,Manton:2022fcb}. 
Thus there is an integer topological charge $Q$ which is given by 
\begin{align}
	Q
	&\equiv -\frac{1}{24\pi^2}\int\dd[3]{x}\epsilon^{jkl}\mathrm{Tr}\qty(U^\dagger\partial_jUU^\dagger\partial_kUU^\dagger\partial_lU),\ \qty{j,k,l}=\qty{1,2,3}\label{Qtop}
\end{align}
where $\epsilon^{jkl}$ is an antisymmetric tensor with $\epsilon^{123}=1$. 
Since $Q$ is a topological invariant, it is unchanged under continuous deformations of $U$, and is used to classify the skyrmion. 

For the imaginary part of the effective action, up to the leading order, one obtains 
\begin{align}
	\Im S_{\mathrm{eff}}
	&=\frac{n_c}{240\pi^2}\int\dd[5]{x}\epsilon_{\alpha\beta\gamma\delta\epsilon}\mathrm{Tr}\qty(A_\alpha A_\beta A_\gamma A_\delta A_\epsilon),\ \qty{\alpha,\beta,\gamma,\delta,\epsilon}=\qty{1,2,3,4,5}\label{ImS}
\end{align}
where we have introduced the $\mathfrak{su}\qty(N)$-valued current $A_\alpha\equiv iU^\dagger\partial_\alpha U$ and an antisymmetric tensor $\epsilon_{\alpha\beta\gamma\delta\epsilon}$ with $\epsilon_{12345}\equiv 1$. 
The topological charge $Q$~\eqref{Qtop} can then be derived from the Noether current associated with the imaginary part of the effective action $\Im S_{\mathrm{eff}}$, when $N>2$~\cite{Diakonov:1997sj}. 
Using a Lorentz trace formula $\mathrm{Tr}\qty(\gamma^j\gamma^k\gamma^l\gamma^0\gamma_5)=4i\epsilon^{jkl}$, the topological charge $Q$ can also be re-expressed in terms of $U^{\gamma_5}$ as 
\begin{align}
	Q=-\frac{1}{96\pi^2i}\int\dd[3]{x}\mathrm{Tr}\qty[\gamma^j\gamma^k\gamma^l\gamma^0\qty(\partial_jU^{\gamma_5})^\dagger\partial_kU^{\gamma_5}\qty(\partial_lU^{\gamma_5})^\dagger U^{\gamma_5}].\label{Qtop5}
\end{align}
This representation is useful in Sec.~\ref{Comp} and Sec.~\ref{Comp2}.

\begin{figure}[t]
	\centering
		\includegraphics[width=0.5\linewidth]{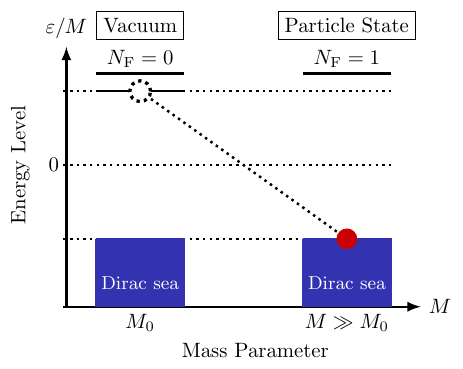}
		\caption{
			A schematic illustration of the fermion-number variation considered in this work. 
			The variation of fermion number $N_{\mathrm{F}}$ between the vacuum and particle states is analyzed using Eq.~\eqref{sfdef} and Eq.~\eqref{sfdef2} with a suitable regularization.
			}
		\label{sf}
\end{figure}

We now return to the CQSM Lagrangian~\eqref{CQSM}. 
For a static chiral matrix $U^{\gamma_5}\qty(x^\mu)=U^{\gamma_5}\qty(x^j)$, the Dirac field can be written in the stationary form $\Psi\qty(x^\mu)=\psi\qty(x^j) e^{-i\varepsilon t}\qty(\varepsilon\in\mathbb{R})$. 
The Dirac equation then becomes an eigenvalue equation for the Dirac Hamiltonian $\mathcal{H}$:
\begin{align}
	\mathcal{H}\psi&=\varepsilon\psi,\ \mathcal{H}\equiv-\gamma^0\qty(i\gamma^j\partial_j-MU^{\gamma_5}),\ j=1,2,3.\label{HCQSM}
\end{align}
Numerous studies have revealed that the fermionic spectrum exhibits characteristic level crossing behavior as the mass parameter $M$ is varied~\cite{Kahana:1984be,Kahana:1984dx,Sawado:1998gk,Sawado:2002zm,Sawado:2004pm}. 
For the standard chiral Yukawa coupling, the net number of such crossings reflects the topological charge $Q$~\eqref{Qtop} of the chiral soliton background. 
The associated spectral rearrangement is accompanied by a change in the induced fermion number, as schematically illustrated in Fig.~\ref{sf}. 
Accordingly, in the next section, we formulate this fermion number in terms of the spectral asymmetry of the Dirac Hamiltonian and define its variation with the mass parameter $M$. 
Note that throughout this paper, we keep the chiral field $U\qty(x)$ fixed as an external background.

\section{Regularized Fermion Number and Its Variation}
\label{General}

Here, we formulate the induced fermion number in terms of the spectral asymmetry of the Dirac Hamiltonian and define its variation with the mass parameter $M$. 
Since the color degrees of freedom produce only a trivial $n_c$-fold degeneracy in the spectrum, we normalize all spectral quantities per color and suppress the overall factor $n_c$ throughout the following analyses. 
The physical fermion number is recovered by multiplying the final result by $n_c$. 
For a given Dirac Hamiltonian $\mathcal{H}$, the fermion number associated with the filled negative-energy states is formally expressed as
\begin{align}
	N_{\mathrm{F}}=\mathrm{Sp}\qty[\theta\qty(-\mathcal{H})],\label{naiveNf}
\end{align}
where $\theta$ denotes the Heaviside step function and $\mathrm{Sp}$ represents the functional trace, i.e., the trace over internal indices (spin, flavor) with integration over physical and momentum spaces~\cite{Diakonov:1997sj}. 
Thus, $N_{\mathrm{F}}$ includes the number of negative energy levels. 
However, $N_{\mathrm{F}}$ is naively divergent when all Dirac-sea contributions are included and therefore requires appropriate regularization. 

To regularize the fermion number $N_{\mathrm{F}}$, we introduce the $\eta$-function
\begin{align}
	\eta\qty(s;\mathcal{H})&\equiv \sum_{\varepsilon\neq0}\mathrm{sign}\qty(\varepsilon)\abs{\varepsilon}^{-s},
\end{align}
which is initially defined for sufficiently large $\Re s$ and analytically continued to $s=0$~\cite{Niemi:1984vz}. 
The $\eta$-invariant is then defined by
\begin{align}
	\eta\qty(\mathcal{H})&\equiv\lim_{s\to0^+}\eta\qty(s;\mathcal{H}).\label{eta_def}
\end{align}
This quantity characterizes the spectral asymmetry between the positive- and negative-energy levels in the spectrum. 
Using the identity $\theta(-x)=\frac{1}{2}\qty(1-\mathrm{sign}(x))$, the formal fermion number in Eq.~\eqref{naiveNf} is decomposed into a background-independent divergent constant and a spectral-asymmetry part. 
Since only the difference from the vacuum is physically meaningful, we subtract the vacuum contribution and define the regularized induced fermion number $N_{\mathrm{F}}^{\mathrm{reg}}$ in terms of the $\eta$-invariant as~\cite{Niemi:1984vz}
\begin{subequations}
	\begin{align}
		N_{\mathrm{F}}^{\mathrm{reg}}\qty(s;M)&\equiv -\frac{1}{2}\eta\qty(s;\mathcal{H}),\label{N_fseta}\\
		N_{\mathrm{F}}^{\mathrm{reg}}\qty(M)&\equiv \lim_{s\to 0^+}N_{\mathrm{F}}^{\mathrm{reg}}\qty(s;M).\label{N_feta}
	\end{align}
\end{subequations} 
To emphasize the mass-dependence, we write the fermion number as $N_{\mathrm{F}}^{\mathrm{reg}}\qty(M)$ which represents the spectral asymmetry of the spectrum\footnote{
	Here, the fermion number denotes the induced fermion number associated with the Dirac sea. 
	The full fermion-number operator contains occupation-number contributions for particle and antiparticle excitations, as well as possible zero-mode contributions when the Hamiltonian has a nontrivial kernel~\cite{Niemi:1984vz}. 
	These contributions depend on the choice of many-body state and are not determined solely by the one-particle Dirac Hamiltonian. 
	In the present work, we focus on the $\eta$-invariant contribution, which characterizes the induced fermion number associated with the spectral asymmetry. 
	Zero modes occurring at isolated intermediate values of $M$ are accounted for through the variation of the $\eta$-invariant along the parameter path.
	} for a given mass parameter $M$. 

Accordingly, rather than considering the regularized fermion number $N_{\mathrm{F}}^{\mathrm{reg}}\qty(M)$ itself, as in the previous studies~\cite{Diakonov:1997sj,Ball:1986fd}, we consider its variation over the mass parameter space~\eqref{parameterspace}. 
This formulation is motivated by the level-crossing picture\footnote{
	When the variation of the $\eta$-invariant arises only from isolated zero-energy level-crossings, the difference in the regularized fermion number $N_{\mathrm{F}}^{\mathrm{reg}}\qty(M)$ over a finite interval $\qty[M_1,M_2]$, $N_{\mathrm{F}}^{\mathrm{reg}}\qty(M_2)-N_{\mathrm{F}}^{\mathrm{reg}}\qty(M_1)$ includes the number of eigenvalues crossing the zero-energy line during the deformation. 
	} as illustrated in Fig.~\ref{sf}. Thus, in the standard CQSM, the fermion-number variation is directly related to the level-crossing picture. 
We then define the fermion-number variation along the path in Eq.~\eqref{parameterspace} as
\begin{subequations}
	\begin{align}
		\Delta N_{\mathrm{F}}\qty(s)
		&\equiv\int_{M_0}^\infty\dd{M}\pdv{N_{\mathrm{F}}^{\mathrm{reg}}\qty(s;M)}{M},\label{sfdef}\\
		\Delta N_{\mathrm{F}}
		&\equiv\lim_{s\to 0^+}\Delta N_{\mathrm{F}}\qty(s).\label{sfdef2}
	\end{align}
\end{subequations}
Note that we implement the limit $s\to 0^+$ by analytic continuation, keeping the regulator parameter $s$ finite during the $M$-integral and taking the limit only after the integration. 
Using Eq.~\eqref{N_fseta} and Eq.~\eqref{N_feta}, the fermion-number variation can be expressed as 
\begin{align}
	\Delta N_{\mathrm{F}}
	=-\lim_{s\to 0^+}\frac{1}{2}\int_{M_0}^\infty\dd{M}\pdv{\eta\qty(s;\mathcal{H})}{M}.\label{Nfcomp}
\end{align}
The derivative with respect to $M$, $\pdv{\eta\qty(s;\mathcal{H})}{M}$ can be represented as 
\begin{align}
	\pdv{\eta\qty(s;\mathcal{H})}{M}=-\mathrm{Sp}\qty[s\qty(\mathcal{H}^2)^{-\frac{s+1}{2}}\pdv{\mathcal{H}}{M}]\label{deta}
\end{align}
where the zero energy modes are subtracted in $\mathrm{Sp}$. 
Since the Hamiltonians considered here depend linearly on $M$, the squared Hamiltonian in Eq.~\eqref{deta} makes explicit whether the mass parameter $M$ is coupled to the background chiral soliton field. 
This relation is the starting point for evaluating the fermion-number variation for different coupling schemes examined in the following sections.

We finally specify the reference point $M=M_0$ introduced in Eq.~\eqref{parameterspace}. 
We choose $M_0$ such that the spectrum has no spectral asymmetry,
\begin{align}
	\eta\qty(\mathcal H)\big|_{M=M_0}=0,
\end{align}
and assume that the Hamiltonian has no zero-mode at this endpoint. 
It then follows from Eq.~\eqref{N_feta} that $N_{\mathrm{F}}^{\mathrm{reg}}\qty(M_0)=0$, so that the above fermion-number variation is measured relative to this reference point. 
At the upper end of the parameter interval, $M\to\infty$, the positive- and negative-energy continua are separated by a gap of order $2M$. 
We therefore evaluate the fermion-number variation over the interval~\eqref{parameterspace}. 


\section{Fermion-Number Variation in the CQSM}
\label{Comp}

In this section, we evaluate the fermion-number variation for the standard chiral Yukawa coupling in the CQSM. 
As established in Sec.~\ref{General}, this variation is determined by the mass derivative of the $\eta$-function, $\pdv{\eta\qty(s;\mathcal{H})}{M}=-\mathrm{Sp}\qty[s\qty(\mathcal{H}^2)^{-\frac{s+1}{2}}\pdv{\mathcal{H}}{M}]$. 
For the CQSM Hamiltonian in Eq.~\eqref{HCQSM}, the interaction with the chiral field appears in the squared Hamiltonian $\mathcal{H}^2$ through a term $V$, defined explicitly in Eq.~\eqref{VdefCQSM}, containing one spatial derivative of $U$. 
Consequently, the resulting expansion organizes the fermion-number variation by the number of factors of $V$ and by the additional derivative corrections generated from their noncommutativity with a derivative operator. 

With this organization, the origin of the topological contribution also becomes apparent from the gamma-matrix structure. 
Each factor of the interaction term $V$ introduces one spatial gamma matrix together with one derivative of the chiral field $U$. 
In three spatial dimensions, the first contribution that produces the antisymmetric tensor $\epsilon^{jkl}$ arises from three such factors through $\mathrm{Tr}\qty(\gamma^j\gamma^k\gamma^l\gamma^0\gamma_5)=4i\epsilon^{jkl}$. 
Thus, the first possible nonvanishing contribution has precisely the three-derivative structure required for the topological charge of the chiral field. 
The remaining terms include higher-order derivative corrections, and we examine whether any of these terms can contribute to the fermion-number variation. 

The key observation is that the derivative of the $\eta$-function naturally decomposes into two parts,
\begin{align}
	\pdv{\eta\qty(s;\mathcal{H})}{M}=\mathcal{Q}_s\qty(M)+\delta_s\qty(M).\label{decomCQSM}
\end{align}
The $\mathcal{Q}_s\qty(M)$ sector yields the topological charge $Q$ after the $M$-integration and the subsequent $s\to0^+$ limit, while the remaining sector $\delta_s\qty(M)$ is shown to vanish term by term. 
We first derive the decomposition~\eqref{decomCQSM} and then evaluate the two parts separately. 

To evaluate the derivative of the $\eta$-function, we first rewrite the squared Dirac Hamiltonian $\mathcal{H}^2$ as 
\begin{subequations}
	\begin{align}
		\mathcal{H}^2
		&=\mathcal{H}_0^2+V,\\ 
		\mathcal{H}_0^2&\equiv-\partial_j^2+M^2,\ V\equiv iM\gamma^j\partial_jU^{\gamma_5}.\label{VdefCQSM}
	\end{align}
\end{subequations}
Then, we obtain 
\begin{align}
	\pdv{\eta\qty(s;\mathcal{H})}{M}=-\mathrm{Sp}\qty[s\qty(\mathcal{H}_0^2+V)^{-\frac{s+1}{2}}\gamma^0U^{\gamma_5}],\label{etasM_CQSM}
\end{align}
where $\pdv{\mathcal{H}}{M}=\gamma^0U^{\gamma_5}$. 
Since $\mathcal{H}^2=\mathcal{H}_0^2+V$ is a self-adjoint operator with a nonnegative spectrum and the zero-energy modes are subtracted in $\mathrm{Sp}$, its fractional power can be represented by the Mellin transform as
\begin{align}
	\qty(\mathcal{H}_0^2+V)^{-\frac{s+1}{2}}
	=\frac{1}{\Gamma\qty(\frac{s+1}{2})}\int_0^\infty\dd{t}t^{\frac{s-1}{2}}e^{-t\qty(\mathcal{H}_0^2+V)},\label{Mellin_CQSM}
\end{align}
where $\Gamma\qty(p)$ is the gamma function. 
We further expand the exponential operator $e^{-t\qty(\mathcal{H}_0^2+V)}$ using the Dyson series as
\begin{align}
	&e^{-t\qty(\mathcal{H}_0^2+V)}
	=e^{-t\mathcal{H}_0^2}\sum_{n=0}^\infty\qty(-1)^n\int_0^t\dd{\tau_1}\int_0^{\tau_1}\dd{\tau_2}\cdots\int_0^{\tau_{n-1}}\dd{\tau_n}\mathcal{V}\qty(\tau_1)\cdots\mathcal{V}\qty(\tau_n)\label{Dyson_CQSM}
\end{align}
where $\mathcal{V}\qty(\tau)\equiv e^{\tau\mathcal{H}_0^2}Ve^{-\tau\mathcal{H}_0^2}$. 
By substituting the Mellin transform~\eqref{Mellin_CQSM} and the Dyson series~\eqref{Dyson_CQSM} into Eq.~\eqref{etasM_CQSM}, and evaluating the functional trace for an operator $\mathcal{O}$, as $\mathrm{Sp}\qty(\mathcal{O})=\int\frac{\dd[3]{k}}{\qty(2\pi)^3}\int\dd[3]{x}\mathrm{Tr}\qty(e^{-ikx}\mathcal{O}e^{ikx})$, we obtain
\begin{align}
	\pdv{\eta\qty(s;\mathcal{H})}{M}
	&=\frac{-s}{\qty(2\pi)^3\Gamma\qty(\frac{s+1}{2})}\int\dd[3]{k}\int_0^\infty\dd{t}t^{\frac{s-1}{2}}e^{-t\qty(k^2+M^2)}\sum_{n=0}^\infty\qty(-1)^n\nonumber\\
	&\times\int\dd[3]{x}\mathrm{Tr}\qty[\int_0^t\dd{\tau_1}\int_0^{\tau_1}\dd{\tau_2}\cdots\int_0^{\tau_{n-1}}\dd{\tau_n}e^{-t\Delta}\mathcal{V}_\Delta\qty(\tau_1)\cdots\mathcal{V}_\Delta\qty(\tau_n)\gamma^0U^{\gamma_5}],\label{etasMafterSP_CQSM}
\end{align}
where $\mathcal{V}_\Delta\qty(\tau)\equiv e^{\tau\Delta}Ve^{-\tau\Delta}$ and $\Delta\equiv-\partial_j^2-2ik_j\partial_j$. 
The operator $\mathcal{V}_\Delta\qty(\tau)$ captures the effects of the noncommutativity of $\Delta$ with the background chiral soliton field $U$. 
To separate the leading contribution from the derivative corrections, we expand the exponential operator $e^{\pm\tau\Delta}$ in $\mathcal{V}_\Delta\qty(\tau)$ and then obtain
\begin{align}
	\mathcal{V}_\Delta\qty(\tau)=\sum_{m=0}^\infty\frac{\tau^m}{m!}\mathrm{ad}_\Delta^m\qty(V).\label{com_def}
\end{align}
Here $\mathrm{ad}_\Delta^m\qty(V)$ denotes the $m$-th iterated commutator which is defined by 
\begin{align}
	\mathrm{ad}_\Delta^0\qty(V)\equiv V,\ \mathrm{ad}_\Delta^1\qty(V)\equiv \qty[\Delta,V],\ \mathrm{ad}_\Delta^m\qty(V)\equiv \qty[\Delta,\mathrm{ad}_\Delta^{m-1}\qty(V)].
\end{align}
Expanding another exponential operator $e^{-t\Delta}$ in Eq.~\eqref{etasMafterSP_CQSM} as $e^{-t\Delta}=\sum_{l=0}^\infty\frac{\qty(-t)^l}{l!}\Delta^l$, the nested integrals are collected into the coefficient $F_{l,\vb{m}}^{\qty(n)}\qty(t)$ as
\begin{align}
	&\int_0^t\dd{\tau_1}\int_0^{\tau_1}\dd{\tau_2}\cdots\int_0^{\tau_{n-1}}\dd{\tau_n}e^{-t\Delta}\mathcal{V}_\Delta\qty(\tau_1)\cdots\mathcal{V}_\Delta\qty(\tau_n)\gamma^0U^{\gamma_5}\nonumber\\
	&=\frac{t^n}{n!}V^n\gamma^0U^{\gamma_5}
	+\sum_{l+\sum_{b=1}^nm_b>0}F_{l,\vb{m}}^{\qty(n)}\qty(t)\Delta^l\prod_{b=1}^{n}\mathrm{ad}_\Delta^{m_b}\qty(V)\gamma^0U^{\gamma_5},\label{nest_CQSM}
\end{align}
where $F_{l,\vb{m}}^{\qty(n)}\qty(t)$ is defined as 
\begin{align}
	F_{l,\vb{m}}^{\qty(n)}\qty(t)
	&\equiv\frac{\qty(-t)^l}{l!}\int_0^t\dd{\tau_1}\int_0^{\tau_1}\dd{\tau_2}\cdots\int_0^{\tau_{n-1}}\dd{\tau_n}\prod_{b=1}^{n}\frac{\tau_b^{m_b}}{m_b!}\nonumber\\
	&=\frac{\qty(-1)^l}{l!}\frac{t^{l+\sum_{b=1}^nm_b+n}}{\prod_{b=1}^nm_b!\prod_{d=1}^n\qty(\sum_{e=d}^nm_e+n-d+1)},
\end{align}
where $\vb{m}\equiv\qty(m_1,\cdots,m_n)$. 
In Eq.~\eqref{nest_CQSM}, we have isolated the commutator-free term, corresponding to $l=m_1=\cdots=m_n=0$, from the remaining terms.
The first term is obtained by replacing all $\mathcal V_\Delta\qty(\tau_b)$ with $V$, with the factor $\frac{t^n}{n!}$ arising from $F_{0,\vb{0}}^{\qty(n)}$. 

Substituting Eq.~\eqref{nest_CQSM} into Eq.~\eqref{etasMafterSP_CQSM}, the derivative of the $\eta$-function is naturally decomposed as 
\begin{align}
	\pdv{\eta\qty(s;\mathcal{H})}{M}
	=\mathcal{Q}_s\qty(M)+\delta_s\qty(M).
\end{align}
The first term consists solely of the leading contribution $V^n$, while the second term contains all remaining corrections generated by $\mathrm{ad}_\Delta^{m_b}\qty(V)$ in Eq.~\eqref{nest_CQSM}. 
The explicit forms of $\mathcal{Q}_s\qty(M)$ and $\delta_s\qty(M)$ are given by
\begin{subequations}
\begin{align}
	\mathcal{Q}_s\qty(M)
	&\equiv 
	\sum_{n=0}^\infty\frac{\qty(-1)^{n+1}\qty(iM)^ns}{n!\qty(2\pi)^3\Gamma\qty(\frac{s+1}{2})}\int\dd[3]{k}\int_0^\infty\dd{t}t^{\frac{s-1}{2}+n}e^{-t\qty(k^2+M^2)}\int\dd[3]{x}\mathrm{Tr}\qty[\qty(\gamma^j\partial_jU^{\gamma_5})^n\gamma^0U^{\gamma_5}],\\
	\delta_s\qty(M)
	&\equiv 
	\sum_{n=0}^\infty\frac{\qty(-1)^{n+1}s}{\qty(2\pi)^3\Gamma\qty(\frac{s+1}{2})}\int\dd[3]{k}\int_0^\infty\dd{t}t^{\frac{s-1}{2}}e^{-t\qty(k^2+M^2)}\nonumber\\
	&\times\int\dd[3]{x}\mathrm{Tr}\qty[\sum_{l+\sum_{b=1}^nm_b>0}F_{l,\vb{m}}^{\qty(n)}\qty(t)\Delta^l\prod_{b=1}^{n}\mathrm{ad}_\Delta^{m_b}\qty(V)\gamma^0U^{\gamma_5}].
\end{align}
\end{subequations}

Next, we summarize the qualitative structure of the calculation. 
In the commutator-free sector $\mathcal{Q}_s\qty(M)$, the terms containing fewer than three factors of $V$ vanish upon taking the Lorentz trace, whereas the $n=3$ term produces the antisymmetric three-derivative structure and yields the topological charge $Q$. 
The terms with $n>3$ in $\mathcal{Q}_s\qty(M)$ do not survive in the $s\to0^+$ limit, as shown by term-by-term analysis. 
The commutator sector $\delta_s\qty(M)$ contains further derivative corrections. 
For each term, to survive in the $s\to0^+$ limit, it must have a specific momentum degree, whereas the commutator structure imposes an upper bound on the available powers of momentum. 
These two conditions are incompatible; consequently, each such contribution vanishes term by term. 
We treat this calculation as a term-by-term expansion, consistent with similar order-by-order analyses of induced fermion-number currents reported in~\cite{Ball:1986fd}. 
We adopt the same approach here. 
In Sec.~\ref{finite_CQSM}, we analyze the commutator-free sector $\mathcal{Q}_s\qty(M)$, and in Sec.~\ref{vanishing_CQSM}, we examine the commutator sector $\delta_s\qty(M)$.

\subsection{Finite part: $\mathcal{Q}_s\qty(M)$}
\label{finite_CQSM}
In this subsection, we evaluate the contribution from $\mathcal{Q}_s\qty(M)$. 
As anticipated, the first nonvanishing contribution in this sector appears with three factors of $V$ and reproduces the topological charge $Q$. 
We first identify this contribution and then examine the remaining terms.

Using an integral formula (see Appendix~\ref{intformulae})
\begin{align}
	&\int\dd[3]{k}\int_0^\infty\dd{t}t^{\frac{s-1}{2}+n}e^{-t\qty(k^2+M^2)}
	=\pi^{\frac{3}{2}}\Gamma\qty(\frac{s-2}{2}+n)M^{-\qty(s-2+2n)},
\end{align}
we can express $\mathcal{Q}_s\qty(M)$ as
\begin{subequations}
\begin{align}
	\mathcal{Q}_s\qty(M)
	&=\sum_{n=0}^\infty\mathcal{Q}_s^{\qty(n)}\qty(M),\\
	\mathcal{Q}_s^{\qty(n)}\qty(M)
	&\equiv\frac{\pi^\frac{3}{2}\qty(-1)^{n+1}i^ns\Gamma\qty(\frac{s-2}{2}+n)}{n!\qty(2\pi)^3\Gamma\qty(\frac{s+1}{2})}M^{-\qty(s-2+n)}\int\dd[3]{x}\mathrm{Tr}\qty[\qty(\gamma^j\partial_jU^{\gamma_5})^n\gamma^0U^{\gamma_5}].\label{etan}
\end{align}
\end{subequations}
Here the index $n$ counts the number of factors of $V=iM\gamma^j\partial_jU^{\gamma_5}$ in the commutator-free sector $\mathcal{Q}_s\qty(M)$. 
We first examine the three lowest orders, which vanish identically owing to the Lorentz trace identities,
\begin{subequations}
\begin{align}
	\mathcal{Q}_s^{\qty(0)}\qty(M)
	&=\frac{\pi^\frac{3}{2}\qty(-1)s\Gamma\qty(\frac{s-2}{2})}{\qty(2\pi)^3\Gamma\qty(\frac{s+1}{2})}M^{-\qty(s-2)}\int\dd[3]{x}\mathrm{Tr}\qty[\gamma^0U^{\gamma_5}]
	=0,
\end{align}
\begin{align}
	\mathcal{Q}_s^{\qty(1)}\qty(M)
	&=\frac{\pi^\frac{3}{2}\qty(-1)^{2}is\Gamma\qty(\frac{s}{2})}{1!\qty(2\pi)^3\Gamma\qty(\frac{s+1}{2})}M^{-\qty(s-1)}\int\dd[3]{x}\mathrm{Tr}\qty[\gamma^j\gamma^0\qty(\partial_jU^{\gamma_5})^\dagger U^{\gamma_5}]
	=0,
\end{align}
\begin{align}
	\mathcal{Q}_s^{\qty(2)}\qty(M)
	&=\frac{\pi^\frac{3}{2}\qty(-1)^{3}i^2s\Gamma\qty(\frac{s+2}{2})}{2!\qty(2\pi)^3\Gamma\qty(\frac{s+1}{2})}M^{-s}\int\dd[3]{x}\mathrm{Tr}\qty[\gamma^j\gamma^k\gamma^0\partial_jU^{\gamma_5}\qty(\partial_kU^{\gamma_5})^\dagger U^{\gamma_5}]
	=0,
\end{align}
\end{subequations}
where we have used the following gamma-matrix formulae:
\begin{align}
	\mathrm{Tr}\qty(\gamma^0)=0,\ \mathrm{Tr}\qty(\gamma^0\gamma_5)=0,\ 
	\mathrm{Tr}\qty(\gamma^j\gamma^0)=0,\nonumber\\ 
	\mathrm{Tr}\qty(\gamma^j\gamma^0\gamma_5)=0,\ 
	\mathrm{Tr}\qty(\gamma^j\gamma^k\gamma^0)=0,\ \mathrm{Tr}\qty(\gamma^j\gamma^k\gamma^0\gamma_5)=0.\label{gamma_CQSM}
\end{align}
The first nonvanishing contribution appears at order $n=3$: 
\begin{align}
	\mathcal{Q}_s^{\qty(3)}\qty(M)
	&=\frac{\pi^\frac{3}{2}\qty(-1)^{4}i^3s\Gamma\qty(\frac{s+4}{2})}{3!\qty(2\pi)^3\Gamma\qty(\frac{s+1}{2})}M^{-\qty(s+1)}\int\dd[3]{x}\mathrm{Tr}\qty[\gamma^j\gamma^k\gamma^l\gamma^0\qty(\partial_jU^{\gamma_5})^\dagger\partial_kU^{\gamma_5}\qty(\partial_lU^{\gamma_5})^\dagger U^{\gamma_5}]\nonumber\\
	&=-2\sqrt{\pi}\frac{s\Gamma\qty(\frac{s+4}{2})}{\Gamma\qty(\frac{s+1}{2})}M^{-\qty(s+1)}Q.
\end{align} 
Here, we have used the expression for the topological charge $Q$ in Eq.~\eqref{Qtop5}. 
As anticipated from the gamma-matrix structure, each of the three factors of $V$ supplies one spatial derivative of the chiral field $\partial_jU$, while the trace over the three spatial gamma matrices produces the antisymmetric tensor $\epsilon^{jkl}$. 
Thus, the topological charge $Q$ appears precisely in the $n=3$ contribution. 
The corresponding contribution to the fermion-number variation is 
\begin{align}
	\Delta N_{\mathrm{F}}^{\qty(3)}
	&\equiv -\lim_{s\to 0^+}\frac{1}{2}\int_{M_0}^\infty\dd{M}\mathcal{Q}_s^{\qty(3)}\qty(M)
	=\lim_{s\to 0^+}\sqrt{\pi}\frac{\Gamma\qty(\frac{s+4}{2})}{\Gamma\qty(\frac{s+1}{2})}M_0^{-s}Q
	=Q.
\end{align}
Thus, the $n=3$ term contributes the topological charge $Q$ to the fermion-number variation.

We next consider the higher-order terms with $n>3$. 
For each fixed $n$, we define the corresponding contribution to the fermion-number variation as 
\begin{align*}
	\Delta N_{\mathrm{F}}^{\qty(n)}\equiv-\lim_{s\to 0^+}\frac{1}{2}\int_{M_0}^\infty\dd{M}\mathcal{Q}_s^{\qty(n)}\qty(M),
\end{align*}
which yields 
\begin{align}
	\Delta N_{\mathrm{F}}^{\qty(n)}
	&=-\lim_{s\to 0^+}\frac{1}{2}\frac{\pi^\frac{3}{2}\qty(-1)^{n+1}i^n\Gamma\qty(\frac{s-2}{2}+n)}{n!\qty(2\pi)^3\Gamma\qty(\frac{s+1}{2})}\frac{sM_0^{-\qty(s-3+n)}}{s-3+n}\int\dd[3]{x}\mathrm{Tr}\qty[\qty(\gamma^j\partial_jU^{\gamma_5})^n\gamma^0U^{\gamma_5}].
\end{align}
For every fixed $n\qty(>3)$, the gamma functions in this expression are finite in the limit $s\to0^+$, and the factor $s-3+n$ in the denominator remains nonzero. 
The explicit factor of $s$ remains uncompensated; therefore,
\begin{align*}
	\Delta N_{\mathrm{F}}^{\qty(n)}=0,\ n>3.
\end{align*}
As indicated above, each higher-derivative contribution in the commutator-free sector vanishes after the $M$-integration and the subsequent $s\to0^+$ limit. 
From these results, the terms with $n=0,1,2$ vanish owing to the Lorentz trace identities, the $n=3$ term yields the topological charge $Q$, and every fixed-order contribution with $n>3$ vanishes in the limit $s\to 0^+$. 
Accordingly, within the term-by-term treatment of the expansion, we obtain
\begin{align}
	\Delta N_{\mathrm{F}}
	&=Q-\lim_{s\to 0^+}\frac{1}{2}\int_{M_0}^\infty\dd{M}\delta_s\qty(M).
\end{align}
It remains to examine the derivative corrections in commutator sector $\delta_s\qty(M)$. 
In the next subsection, we show that each contribution in this sector vanishes, term by term, after the $M$-integration and the limit $s\to0^+$. 

\subsection{Vanishing part: $\delta_s\qty(M)$}
\label{vanishing_CQSM}
In this subsection, we examine the remaining derivative corrections contained in $\delta_s\qty(M)$. 
This sector contains terms with at least one power of $\Delta$ or at least one commutator $\qty[\Delta,V]$. 
We show that each such contribution vanishes after the $M$-integration and the subsequent $s\to0^+$ limit. 
To this end, we decompose $\delta_s\qty(M)$ according to $n$, the number of $V$, and the additional indices $\qty(l,\vb{m})$ which represent the number of $\Delta$ or $\qty[\Delta,V]$ as
\begin{subequations}
\begin{align}
	\delta_s\qty(M)
	&=\sum_{n=0}^\infty\sum_{l+\sum_{b=1}^nm_b>0}\delta_{s,l,\vb{m}}^{\qty(n)}\qty(M),\\
	\delta_{s,l,\vb{m}}^{\qty(n)}\qty(M)
	&\equiv 
	\frac{\qty(-1)^{n+1}s}{\qty(2\pi)^3\Gamma\qty(\frac{s+1}{2})}\int\dd[3]{k}\int_0^\infty\dd{t}t^{\frac{s-1}{2}}e^{-t\qty(k^2+M^2)}\int\dd[3]{x}\mathrm{Tr}\qty[F_{l,\vb{m}}^{\qty(n)}\qty(t)\Delta^l\prod_{b=1}^{n}\mathrm{ad}_\Delta^{m_b}\qty(V)\gamma^0U^{\gamma_5}].
\end{align}
\end{subequations}

We first examine the lowest orders explicitly. 
The three lowest orders vanish identically owing to the Lorentz trace identities,
\begin{subequations}
\begin{align}
	&\delta_{s,l,\vb{m}}^{\qty(0)}\qty(M)\nonumber\\
	&=
	\frac{\qty(-1)s}{\qty(2\pi)^3\Gamma\qty(\frac{s+1}{2})}\int\dd[3]{k}\int_0^\infty\dd{t}t^{\frac{s-1}{2}}e^{-t\qty(k^2+M^2)}\int\dd[3]{x}\mathrm{Tr}\qty[\frac{\qty(-t)^l}{l!}\Delta^l\gamma^0U^{\gamma_5}]\nonumber\\
	&=0,
\end{align}
\begin{align}
	&\delta_{s,l,\vb{m}}^{\qty(1)}\qty(M)\nonumber\\
	&=
	\frac{s}{\qty(2\pi)^3\Gamma\qty(\frac{s+1}{2})}\int\dd[3]{k}\int_0^\infty\dd{t}t^{\frac{s-1}{2}}e^{-t\qty(k^2+M^2)}\int\dd[3]{x}\mathrm{Tr}\qty[F_{l,\vb{m}}^{\qty(1)}\qty(t)\Delta^l\mathrm{ad}_\Delta^{m_1}\qty(V)\gamma^0U^{\gamma_5}]\nonumber\\
	&=0,
\end{align}
\begin{align}
	&\delta_{s,l,\vb{m}}^{\qty(2)}\qty(M)\nonumber\\
	&=
	\frac{\qty(-1)s}{\qty(2\pi)^3\Gamma\qty(\frac{s+1}{2})}\int\dd[3]{k}\int_0^\infty\dd{t}t^{\frac{s-1}{2}}e^{-t\qty(k^2+M^2)}\int\dd[3]{x}\mathrm{Tr}\qty[F_{l,\vb{m}}^{\qty(2)}\qty(t)\Delta^l\prod_{b=1}^{2}\mathrm{ad}_\Delta^{m_b}\qty(V)\gamma^0U^{\gamma_5}]\nonumber\\
	&=0,
\end{align}
\end{subequations}
where we have used the trace formulae of the gamma matrices~\eqref{gamma_CQSM}. 
Since $\Delta$ carries no Lorentz indices, $\Delta^l$ and $\mathrm{ad_\Delta^{m_b}}$ do not alter the gamma-matrix structure in these vanishing traces.

For $n\geq3$, however, the Lorentz trace alone does not eliminate the remaining terms. 
We therefore derive a necessary condition for a contribution $\delta^{\qty(n)}_{s,l,\vb{m}}\qty(M)$ to survive the $M$-integration and subsequent limit $s\to0^+$. 
To derive this condition, we first perform the momentum integral together with the integral arising from the Mellin transform~\eqref{Mellin_CQSM} in $\delta_{s,l,\vb{m}}^{\qty(n)}\qty(M)$. 
Considering $F_{l,\vb{m}}^{\qty(n)}\propto t^{l+m+n}$ with $m\equiv \sum_{b=1}^nm_b$ and $\prod_{b=1}^n\mathrm{ad}_\Delta^{m_b}\qty(V)\propto M^n$, and further assuming that a term in the spatial integral contains $k_1^{c_1}k_2^{c_2}k_3^{c_3}\ \qty(c_i\in\mathbb{Z}_{\geq0})$, we obtain 
\begin{align}
	&M^n\int\dd[3]{k}\int_0^\infty\dd{t}t^{\frac{s-1}{2}}e^{-t\qty(k^2+M^2)}t^{l+m+n}k_1^{c_1}k_2^{c_2}k_3^{c_3}\nonumber\\
	&=\mathcal{G}\qty(c_1,c_2,c_3)\Gamma\qty(\frac{s-2-c+2l+2m}{2}+n)M^{-\qty(s-2-c+2l+2m+n)},
\end{align}
where $\mathcal{G}\qty(c_1,c_2,c_3)\equiv\prod_{i=1}^3\frac{1+\qty(-1)^{c_i}}{2}\Gamma\qty(\frac{1+c_i}{2})$ and $c\equiv\sum_{i=1}^3c_i$ (see Appendix~\ref{intformulae}). 
Note that if any $c_i$ is odd, $\mathcal{G}\qty(c_1,c_2,c_3)$ is identically zero; consequently, the corresponding term in $\delta_{s,l,\vb{m}}^{\qty(n)}\qty(M)$ vanishes. 
Thus, for a term to contribute to the fermion-number variation $\Delta N_{\mathrm{F}}$, each $c_i$ must be a nonnegative even integer. 
Integrating $\delta_{s,l,\vb{m}}^{\qty(n)}\qty(M)$ over $M$, we obtain 
\begin{align}
	&\int_{M_0}^\infty\dd{M}\delta_{s,l,\vb{m}}^{\qty(n)}\qty(M)
	\propto\frac{\Gamma\qty(\frac{s-2-c+2l+2m}{2}+n)}{\Gamma\qty(\frac{s+1}{2})}\frac{sM_0^{-\qty(s-3-c+2l+2m+n)}}{s-3-c+2l+2m+n}.\label{intM}
\end{align}
For this term to survive in the $s\to 0^+$ limit, the explicit factor of $s$ must be compensated by a singular factor $s^{-1}$. 
As shown below, the momentum-degree bound $c\le l+m$ guarantees that the gamma function in Eq.~\eqref{intM} is finite at $s=0$ for $n\ge3$. 
Therefore, a nonvanishing contribution requires the denominator generated by the $M$-integration to vanish at $s=0$, yielding
\begin{align}
	2l+2m+n-3=c.
\end{align}
This condition is necessary for $\delta_{s,l,\vb{m}}^{\qty(n)}\qty(M)$ to contribute to the fermion-number variation $\Delta N_{\mathrm{F}}$. 
Since $c$ is a nonnegative even integer, the order index $n$ must be a positive odd integer to satisfy the necessary condition. 
Thus, for $n\geq 3$, we use $n=2p+3\ \qty(p\in\mathbb{Z}_{\geq 0})$ and obtain
\begin{align}
	2l+2m+2p=c.\label{deltafinite}
\end{align}
It remains to determine whether this condition can actually be satisfied.

To this end, we analyze the momentum dependence of each term in $\delta_{s,l,\vb{m}}^{\qty(n)}\qty(M)$ to determine the range of possible values of $c$. 
Note that the momentum dependence arises solely from $\Delta^l$ and $\mathrm{ad}_\Delta^{m_b}\qty(V)$. 
The operator $\Delta^l$ can be expanded as 
\begin{align}
	\Delta^l
	&=\qty(-1)^l\sum_{r=0}^l\sum_{u=0}^r\sum_{v=0}^u
	\binom{l}{r}\binom{r}{u}\binom{u}{v}
	\qty(2i)^r\qty(\partial_j^2)^{l-r}\partial_1^{r-u}\partial_2^{u-v}\partial_3^{v}k_1^{r-u}k_2^{u-v}k_3^v.
\end{align}
Hence, each term in the expansion has momentum dependence of the form
\begin{align}
	k_1^{r-u}k_2^{u-v}k_3^v\ \qty(0\leq r\leq l,\ 0\leq u\leq r,\ 0\leq v\leq u). \label{Deltaexpansion}
\end{align}
By expanding $\Delta$ in the same manner, we apply an analogous argument to $\mathrm{ad}_\Delta^{m_b}\qty(V)$ $\qty(b=1,\cdots,n)$ and obtain the momentum dependence
\begin{align}
	k_1^{r_b-u_b}k_2^{u_b-v_b}k_3^{v_b}\ \qty(0\leq r_b\leq m_b,\ 0\leq u_b\leq r_b,\ 0\leq v_b\leq u_b). \label{ad_Delta^m_bexpansion}
\end{align}
Combining these factors, we find that every term in $\delta_{s,l,\vb{m}}^{\qty(n)}\qty(M)$ has the momentum dependence of the form 
\begin{align}
	k_1^{r-u}k_2^{u-v}k_3^{v}k_1^{r_1-u_1}k_2^{u_1-v_1}k_3^{v_1}\cdots k_1^{r_n-u_n}k_2^{u_n-v_n}k_3^{v_n}.
\end{align}
Therefore, the corresponding momentum index $c$ becomes
\begin{align}
	c
	&=r+\sum_{b=1}^nr_b.
\end{align}
Since $r$ and $r_b$ satisfy $0\leq r\leq l$ and $0\leq r_b\leq m_b$ respectively, one immediately finds
\begin{align}
	0\leq c\leq l+m.
\end{align}
Note that this bound also ensures
\begin{align}
	\frac{-2-c+2l+2m}{2}+n\geq \frac{-2+l+m}{2}+n>0,
\end{align}
confirming that the gamma function in Eq.~\eqref{intM} is regular at $s=0$. 
This bound is, however, incompatible with the necessary condition~\eqref{deltafinite}, because 
\begin{align}
	2l+2m+2p\geq 2l+2m> l+m\geq c.
\end{align}
Therefore, no term in $\delta_{s}\qty(M)$ contributes to the fermion-number variation.

Hence, for every allowed set of indices $\qty(n,l,\vb{m})$ in $\delta_s\qty(M)$, we conclude that
\begin{align}
	\lim_{s\to 0^+}\frac{1}{2}\int_{M_0}^\infty\dd{M}\delta^{\qty(n)}_{s,l,\vb{m}}\qty(M)=0.
\end{align}
Thus, every term in $\delta_s\qty(M)$ vanishes after the $M$-integration and the subsequent $s\to0^+$ limit. 
Combining this result with the commutator-free contribution derived in Sec.~\ref{finite_CQSM}, we find that, within the term-by-term treatment, 
\begin{align}
	\Delta N_{\mathrm{F}}=Q
\end{align}
For the standard CQSM, this result is consistent with the level-crossing picture in which the net rearrangement of the fermionic levels reflects the topological charge of the chiral soliton background.


\section{Two alternative couplings}
\label{Comp2}

In Sec.~\ref{Comp}, we have evaluated the fermion-number variation for the standard chiral Yukawa coupling of the CQSM and obtained $\Delta N_{\mathrm{F}}=Q$ within the term-by-term expansion. 
This result raises the question of whether this correspondence is specific to the chiral Yukawa coupling of the CQSM or reflects a more general property of coupling schemes. 
To address this question, we consider two alternative couplings derived from the same chiral field: an axial-vector coupling obtained by an $\mathrm{SU}\qty(N)$ chiral rotation of the standard CQSM and a vector coupling obtained by removing the chiral projection operator from the axial-vector coupling. 
In both cases, the underlying chiral field and its topological charge $Q$ remain unchanged, whereas the coupling structure is modified. 
As shown below, the axial-vector coupling retains the same fermion-number variation as the standard CQSM, $\Delta N_{\mathrm{F}}=Q$, whereas the vector coupling yields $\Delta N_{\mathrm{F}}=0$. 
Thus, the same topologically nontrivial chiral soliton background leads to different fermion-number variations depending on the coupling scheme. 

\subsection{Axial-vector coupling}
\label{Comp21}
In this subsection, we first consider an axial-vector coupling associated with the standard CQSM~\eqref{CQSM}. 
Under the $\mathrm{SU}\qty(N)$ chiral rotation\footnote{This transformation corresponds to $\Psi_+\to\Psi_+,\ \Psi_-\to U\Psi_-$, where $\Psi_\pm\equiv P_\pm\Psi$, and does not change the fermion density $\Psi^\dagger\Psi$}, 
\begin{align}
	\Psi\to\qty(P_++UP_-)\Psi,\label{SU2rot}
\end{align}
with $P_\pm=\frac{1\pm\gamma_5}{2}$, the Lagrangian~\eqref{CQSM} can be reformulated as 
\begin{align}
	\mathcal{L}=\overline{\Psi}\qty[i\gamma^\mu\qty(\partial_\mu-iP_-A_\mu)-M]\Psi,\label{CQSMrot}
\end{align}
where $A_\mu\equiv iU^\dagger\partial_\mu U$ represents the $\mathfrak{su}\qty(N)$-valued current~\cite{BALACHANDRAN:1999}. 
This representation describes an axial-vector coupling between the Dirac and chiral fields. 
In the following, we restrict the chiral field to be static, so that $A_0=0$, and $A_\mu$ has a pure-gauge form; however, the more general case is briefly discussed in Sec.~\ref{Remark}. 

We then evaluate the fermion-number variation for this representation. 
Unlike the standard CQSM, the squared Dirac Hamiltonian contains an additional first-order differential operator $K$ arising from the axial-vector coupling: 
\begin{subequations}
\begin{align}
	\mathcal{H}^2
	=\mathcal{H}_0^2+K+V,\ 
	\mathcal{H}
	=-\gamma^0\qty[i\gamma^j\qty(\partial_j-iP_-A_j)-M],\label{HCQSMrot}
\end{align}
\begin{align}
	\mathcal{H}_0^2\equiv-\partial_j^2+M^2,\ 
	K\equiv2iP_-A_j\partial_j,\ 
	V\equiv-i\gamma^j\gamma^kP_-D_jA_k+M\gamma^j\gamma_5A_j
\end{align}
\end{subequations}
where $D_j\equiv\partial_j-iA_j$ is the covariant derivative. 
The interaction $V$ contains both a mass-independent term and the background-dependent term $M\gamma^j\gamma^5A_j$, which is linear in the mass parameter. 
Besides the additional first-order differential operator $K$, these interaction terms modify the organization of the expansion relative to that in Sec.~\ref{Comp}. 

Before the explicit expansion, we summarize the structure of the calculation. 
In contrast to the standard CQSM, the interaction now contains both mass-independent terms and the mass-dependent term $M\gamma^j\gamma^5 A_j$. 
The latter provides the contribution generating the topological charge. 
For the pure-gauge background considered here, the surviving topological contribution arises from three factors of this mass-dependent term and reproduces the topological charge $Q$, because $A_j$ contains one derivative. 
At lower order, the necessary condition also permits an $n=2$ contribution, but it vanishes for the pure-gauge form $A_j=iU^\dagger\partial_jU$. 
Higher-order terms are excluded term by term, by the degree condition required for a nonvanishing contribution after the $M$-integration and the $s\to0^+$ limit. 
The additional contributions generated by explicit factors of $K$ and the derivative corrections associated with $\Delta$ are eliminated through the degree-counting argument. 

Following the same term-by-term procedure as in Sec.~\ref{Comp}, we evaluate the derivative of the $\eta$-function by employing the Mellin transform and the Dyson series. 
The first-order differential operator $K$ distinguishes this case from the standard CQSM and generates additional terms when the functional trace is evaluated in the plane-wave basis. 
Consequently, besides the contributions included within the standard CQSM, additional terms arise in the expansion of the functional trace. 
Accordingly, the derivative of the $\eta$-function can be decomposed as 
\begin{align}
	\pdv{\eta\qty(s;\mathcal{H})}{M}=\mathcal{Q}_s\qty(M)+\xi_s\qty(M)+\delta_s\qty(M).\label{etanaxial}
\end{align}
The three terms are distinguished by their structure in the expansion. 
After evaluating the functional trace in the plane-wave basis, we introduce $K'\equiv-2P_-A_jk_j$. 
The contribution $\mathcal{Q}_s\qty(M)$ contains only factors from $K'+V$, including the momentum-dependent factor $K'$. 
The term $\xi_s\qty(M)$ includes the remaining contributions containing one or more explicit factors of $K$, while $\delta_s\qty(M)$ includes those containing at least one power of $\Delta$ or one commutator $\qty[\Delta,K'+V+K]$. 
Their explicit forms are 
\begin{subequations}
\begin{align}
	\mathcal{Q}_s\qty(M)
	&\equiv \sum_{n=0}^\infty\frac{\qty(-1)^{n+1}s}{n!\qty(2\pi)^3\Gamma\qty(\frac{s+1}{2})}\int\dd[3]{k}\int_0^\infty\dd{t}t^{\frac{s-1}{2}+n}e^{-t\qty(k^2+M^2)}\nonumber\\
	&\times\int\dd[3]{x}\mathrm{Tr}\qty[\qty(-2P_-A_jk_j-i\gamma^j\gamma^kP_-D_jA_k
	+M\gamma^j\gamma_5A_j)^n\gamma^0],
\end{align}
\begin{align}
	\xi_s\qty(M)
	&\equiv\sum_{n=0}^\infty\frac{\qty(-1)^{n+1}s}{n!\qty(2\pi)^3\Gamma\qty(\frac{s+1}{2})}\int\dd[3]{k}\int_0^\infty\dd{t}t^{\frac{s-1}{2}+n}e^{-t\qty(k^2+M^2)}\nonumber\\
	&\times\int\dd[3]{x}\mathrm{Tr}\qty[\sum_{r=1}^n\sum_{p_0+\cdots+p_r=n-r}\qty(K'+V)^{p_0}\prod_{r'=1}^r\qty[K\qty(K'+V)^{p_{r'}}]\gamma^0],
\end{align}
\begin{align}
	\delta_s\qty(M)
	&\equiv\sum_{n=0}^\infty\frac{\qty(-1)^{n+1}s}{\qty(2\pi)^3\Gamma\qty(\frac{s+1}{2})}\int\dd[3]{k}\int_0^\infty\dd{t}t^{\frac{s-1}{2}}e^{-t\qty(k^2+M^2)}\nonumber\\
	&\times\int\dd[3]{x}\mathrm{Tr}\qty[\sum_{l+\sum_{b=1}^nm_b>0}F_{l,\vb{m}}^{\qty(n)}\qty(t)\Delta^l\prod_{b=1}^n\mathrm{ad}_\Delta^{m_b}\qty(K'+V+K)\gamma^0],
\end{align}
\end{subequations}
where $K'\equiv -2P_-A_jk_j$. 
The analysis in the following subsections is parallel to the analysis of Sec.~\ref{Comp}. 
We show that $\mathcal{Q}_s\qty(M)$ yields the topological charge $Q$, whereas the additional sectors $\xi_s\qty(M)$ and $\delta_s\qty(M)$ do not contribute to the fermion-number variation term by term.

\subsubsection{Finite part: $\mathcal{Q}_s\qty(M)$}
\label{Finite_axial}
Here, we examine the contribution from $Q_s(M)$, which is formally expanded according to the number of factors of $K'+V$ as
\begin{subequations}
\begin{align}
	\mathcal{Q}_s\qty(M)
	&=\sum_{n=0}^\infty\mathcal{Q}_s^{\qty(n)}\qty(M),\\
	\mathcal{Q}_s^{\qty(n)}\qty(M)
	&\equiv\frac{\qty(-1)^{n+1}s}{n!\qty(2\pi)^3\Gamma\qty(\frac{s+1}{2})}\int\dd[3]{k}\int_0^\infty\dd{t}t^{\frac{s-1}{2}+n}e^{-t\qty(k^2+M^2)}\nonumber\\
	&\times\int\dd[3]{x}\mathrm{Tr}\qty[\qty(-2P_-A_jk_j-i\gamma^j\gamma^kP_-D_jA_k
	+M\gamma^j\gamma_5A_j)^n\gamma^0].
\end{align}
\end{subequations}

Unlike the standard CQSM, the first-order differential operator $K$ induces explicit momentum dependence into the spatial integral. 
Moreover, since the interaction contains both mass-dependent and mass-independent terms, the power of $M$ is no longer fixed at a given order $n$. 
Before evaluating each order, we therefore derive a necessary condition for a fixed $n$ term to remain nonzero after the $M$-integration and the subsequent $s\to0^+$ limit. 
This condition identifies the terms that can contribute to the fermion-number variation without explicitly expanding the $n$-th power in the spatial integral. 
Assuming that a term in the spatial integral contains a factor $M^\beta k_1^{a_1}k_2^{a_2}k_3^{a_3}\ \qty(\beta,a_i\in\mathbb{Z}_{\geq 0})$, we perform the integrals over $k$ and $t$ as 
\begin{align}
	&M^\beta\int\dd[3]{k}\int_0^\infty\dd{t}t^{\frac{s-1}{2}+n}e^{-t\qty(k^2+M^2)}k_1^{a_1}k_2^{a_2}k_3^{a_3}\nonumber\\
	&=\mathcal{G}\qty(a_1,a_2,a_3)\Gamma\qty(\frac{s-2-a}{2}+n)M^{-\qty(s-2-a+2n-\beta)}
\end{align}
where $\mathcal{G}\qty(a_1,a_2,a_3)\equiv\prod_{i=1}^3\frac{1+\qty(-1)^{a_i}}{2}\Gamma\qty(\frac{1+a_i}{2})$ and $a\equiv\sum_{i=1}^3a_i$. 
Note that if any $a_i$ is odd, $\mathcal{G}\qty(a_1,a_2,a_3)$ is identically zero; consequently, the corresponding term in $\mathcal{Q}_s^{\qty(n)}\qty(M)$ vanishes. 
Therefore, a nonzero contribution requires every $a_i$ to be a nonnegative even integer. 
For $n>1$, the $M$-integration yields 
\begin{align}
	\int_{M_0}^\infty\dd{M}\mathcal{Q}_s^{\qty(n)}\qty(M)
	\propto \frac{\Gamma\qty(\frac{s-2-a}{2}+n)}{\Gamma\qty(\frac{s+1}{2})}\frac{sM_0^{-\qty(s-3-a+2n-\beta)}}{s-3-a+2n-\beta}.
\end{align}
For a fixed term to survive the limit $s\to0^+$, the explicit factor of $s$ must be compensated by the denominator $s^{-1}$ resulting from the $M$-integration. 
This requirement yields the necessary condition
\begin{align}
	2n-3&=a+\beta\label{finite}.
\end{align}
This condition is necessary for $\mathcal{Q}_s^{\qty(n)}\qty(M)$ to contribute finitely to the fermion-number variation. 
If it is satisfied, the gamma function in $\mathcal{Q}_s^{\qty(n)}\qty(M)$ is finite at $s=0$. 

We begin with the lowest two orders. 
Since the above necessary condition applies only to $n>1$, the cases $n=0$ and $n=1$ must be evaluated directly. 
These terms vanish identically because of the Lorentz trace identities,
\begin{subequations}
\begin{align}
	\mathcal{Q}_s^{\qty(0)}\qty(M)
	&=\frac{\qty(-1)s}{\qty(2\pi)^3\Gamma\qty(\frac{s+1}{2})}\int\dd[3]{k}\int_0^\infty\dd{t}t^{\frac{s-1}{2}}e^{-t\qty(k^2+M^2)}\int\dd[3]{x}\mathrm{Tr}\qty(\gamma^0)\nonumber\\
	&=0,
\end{align}
\begin{align}
	\mathcal{Q}_s^{\qty(1)}\qty(M)
	&=\frac{\qty(-1)^2s}{\qty(2\pi)^3\Gamma\qty(\frac{s+1}{2})}\int\dd[3]{k}\int_0^\infty\dd{t}t^{\frac{s-1}{2}+1}e^{-t\qty(k^2+M^2)}\nonumber\\
	&\times\int\dd[3]{x}\mathrm{Tr}\qty[\qty(-2P_-A_jk_j-i\gamma^j\gamma^kP_-D_jA_k
	+M\gamma^j\gamma_5A_j)\gamma^0]\nonumber\\
	&=0,
\end{align}
\end{subequations}
where we have used the following gamma-matrix formulae:
\begin{align}
	\mathrm{Tr}\qty(\gamma^0)=0,\ 
	\mathrm{Tr}\qty(P_-\gamma^0)=0,\ \mathrm{Tr}\qty(\gamma^j\gamma^kP_-\gamma^0)=0,\ \mathrm{Tr}\qty(\gamma^j\gamma_5\gamma^0)=0.\label{gamma_axial}
\end{align}

We next consider the remaining orders that satisfy the necessary condition. 
For $n=2$, the necessary condition~\eqref{finite} becomes 
\begin{align}
	a+\beta=1.
\end{align}
Since $a$ must be a nonnegative even integer, the only possible solution is
\begin{align}
	a=0,\ \beta=1.\label{solaxin2}
\end{align}
This solution uniquely determines the structure of the surviving term. 
The condition $a=0$ excludes factors of $K'=-2P_-A_jk_j$, whereas $\beta=1$ requires a single factor of the mass-dependent term $M\gamma^j\gamma_5A_j$. 
Consequently, the spatial integral reduces to
\begin{align}
	\int\dd[3]{x}\mathrm{Tr}\qty[\qty(-i\gamma^j\gamma^kP_-\gamma^l\gamma_5D_jA_kA_l-i\gamma^j\gamma_5\gamma^k\gamma^lP_-A_jD_kA_l)\gamma^0].
\end{align}
Further, using the relations
\begin{align}
	\mathrm{Tr}\qty(\gamma^j\gamma^k\gamma^l\gamma^0\gamma_5)=4i\epsilon^{jkl},\ \mathrm{Tr}\qty(\gamma^j\gamma^k\gamma^l\gamma^0)=0,
\end{align}
the spatial integral becomes
\begin{align}
	-4\int\dd[3]{x}\epsilon^{jkl}\mathrm{Tr}\qty(D_jA_kA_l).
\end{align}
Substituting the pure-gauge form $A_j=iU^\dagger\partial_jU$ together with $U^\dagger U=1$, one finds that the above integral vanishes. 
Therefore, the second-order term $\qty(n=2)$ does not contribute to the fermion-number variation.

For $n=3$, the necessary condition~\eqref{finite} becomes 
\begin{align}
	a+\beta=3.
\end{align}
This condition admits two possible combinations:
\begin{align}
	\qty(a,\beta)=\qty(0,3),\qty(2,1).
\end{align}
We first consider the case of $\qty(a,\beta)=\qty(2,1)$. 
In this case, the corresponding terms contain two factors of $K'=-2P_-A_jk_j$ and one factor of $M\gamma^j\gamma_5A_j$. 
However, the corresponding Lorentz trace vanishes, 
\begin{align}
	&\mathrm{Tr}\qty[\qty(-2P_-A_jk_j)\gamma^k\gamma_5A_k\qty(-2P_-A_lk_l)\gamma^0]\nonumber\\
	&+\mathrm{Tr}\qty[\gamma^j\gamma_5A_j\qty(-2P_-A_kk_k)^2\gamma^0]+\mathrm{Tr}\qty[\qty(-2P_-A_jk_j)^2\gamma^k\gamma_5A_k\gamma^0]\nonumber\\
	&=4\mathrm{Tr}\qty[\gamma^l\gamma_5\gamma^0A_jA_kA_lk_jk_k]\nonumber\\
	&=0,
\end{align}
where we have used $\mathrm{Tr}\qty(\gamma^l\gamma_5\gamma^0)=0$. 
The remaining possibility is therefore $\qty(a,\beta)=\qty(0,3)$. 
In this case, the spatial integral reduces to
\begin{align}
	\int\dd[3]{x}\mathrm{Tr}\qty[\qty(\gamma^j\gamma_5A_j)^3\gamma^0]=-96\pi^2Q
\end{align}
where the representation of $Q$~\eqref{Qtop5} is used. 
Consequently, the third-order contribution to the fermion-number variation becomes
\begin{align}
	\Delta N_{\mathrm{F}}^{\qty(3)}
	&\equiv-\frac{1}{2}\lim_{s\to 0^+}\int_{M_0}^\infty\dd{M}\mathcal{Q}_s^{\qty(3)}\qty(M)
	=\lim_{s\to 0^+}\frac{48\pi^2Q\pi^{\frac{3}{2}}\Gamma\qty(\frac{s+4}{2})}{3!\qty(2\pi)^3\Gamma\qty(\frac{s+1}{2})}M_0^{-s}
	=Q.
\end{align}
Thus, the finite contribution arises at $n=3$. 

Finally, we consider the higher-order terms with $n>3$. 
The necessary condition~\eqref{finite} requires $a+\beta=2n-3$. 
By construction, however, the total degree of $k_j$ and $M$ in an $n$-th order term cannot exceed $n$, that is
\begin{align}
	a+\beta\leq n.
\end{align}
Further combining these two relations, we obtain 
\begin{align}
	a+\beta=2n-3\leq n
\end{align}
which is equivalent to 
\begin{align}
	n\leq 3.
\end{align}
Hence, the necessary condition cannot be satisfied for $n>3$. 
Therefore, all higher-order terms vanish term by term after the $M$-integration and the subsequent $s\to 0^+$ limit, and $\mathcal{Q}_s^{\qty(n)}\qty(M)$ yields no finite contribution for $n>3$.

Collecting the above results, within the term-by-term analysis, we obtain
\begin{align}
	\Delta N_{\mathrm{F}}
	&=Q-\lim_{s\to 0^+}\frac{1}{2}\int_{M_0}^\infty\dd{M}\qty(\xi_s\qty(M)+\delta_s\qty(M)).
\end{align}
Thus the remaining task is to analyze the contributions from $\xi_s(M)$ and $\delta_s(M)$, as detailed in the following subsection.

\subsubsection{Vanishing part: $\xi_s\qty(M)$, $\delta_s\qty(M)$}
\label{Vanish_axial}
In this subsection, we analyze the remaining contribution to the fermion-number variation: 
\begin{align}
	\lim_{s\to 0^+}\frac{1}{2}\int_{M_0}^\infty\dd{M}\qty(\xi_s\qty(M)+\delta_s\qty(M)).
\end{align}
We first examine the terms originating from the additional operator $K$, namely $\xi_s\qty(M)$, and then analyze $\delta_s\qty(M)$. 
For later convenience, we decompose $\xi_s\qty(M)$ and $\delta_s\qty(M)$ into their $n$-th order contributions and further resolve them in terms of their remaining indices as 
\begin{subequations}
\begin{align}
	\xi_s\qty(M)
	&=\sum_{n=0}^\infty\sum_{r=1}^n\xi_{s,r}^{\qty(n)}\qty(M),\nonumber
\end{align}
\begin{align}
	&\xi_{s,r}^{\qty(n)}\qty(M)\equiv\frac{\qty(-1)^{n+1}s}{n!\qty(2\pi)^3\Gamma\qty(\frac{s+1}{2})}\int\dd[3]{k}\int_0^\infty\dd{t}t^{\frac{s-1}{2}+n}e^{-t\qty(k^2+M^2)}\nonumber\\
	&\times\int\dd[3]{x}\mathrm{Tr}\qty[\sum_{p_0+\cdots+p_r=n-r}\qty(K'+V)^{p_0}\prod_{r'=1}^r\qty[K\qty(K'+V)^{p_{r'}}]\gamma^0],
\end{align}
\begin{align}
	\delta_s\qty(M)=\sum_{n=0}^\infty\sum_{l+\sum_{b=1}^nm_b>0}\delta_{s,l,\vb{m}}^{\qty(n)}\qty(M),\nonumber
\end{align}
\begin{align}
	&\delta_{s,l,\vb{m}}^{\qty(n)}\qty(M)\equiv\frac{\qty(-1)^{n+1}s}{\qty(2\pi)^3\Gamma\qty(\frac{s+1}{2})}\int\dd[3]{k}\int_0^\infty\dd{t}t^{\frac{s-1}{2}}e^{-t\qty(k^2+M^2)}\nonumber\\
	&\times\int\dd[3]{x}\mathrm{Tr}\qty[F_{l,\vb{m}}^{\qty(n)}\qty(t)\Delta^l\prod_{b=1}^n\mathrm{ad}_\Delta^{m_b}\qty(K'+V+K)\gamma^0].
\end{align}
\end{subequations}

We first focus on the contributions from $\xi_s\qty(M)$. 
Here, the $n=0$ contribution is absent because the sum over $r$ is empty. 
For the next two orders, $n=1$ and $n=2$, we obtain
\begin{align}
	\xi_{s,1}^{\qty(1)}\qty(M)
	&=\frac{s}{\qty(2\pi)^3\Gamma\qty(\frac{s+1}{2})}\int\dd[3]{k}\int_0^\infty\dd{t}t^{\frac{s-1}{2}+1}e^{-t\qty(k^2+M^2)}\int\dd[3]{x}\mathrm{Tr}\qty[K\gamma^0]\nonumber\\
	&=0,
\end{align}
\begin{align}
	\sum_{r=1}^2\xi_{s,r}^{\qty(2)}\qty(M)
	&=\frac{\qty(-1)s}{2!\qty(2\pi)^3\Gamma\qty(\frac{s+1}{2})}\int\dd[3]{k}\int_0^\infty\dd{t}t^{\frac{s-1}{2}+2}e^{-t\qty(k^2+M^2)}\nonumber\\
	&\times\int\dd[3]{x}\mathrm{Tr}\qty[\qty(K'+V)K\gamma^0+K\qty(K'+V)\gamma^0+K^2\gamma^0],\nonumber\\
	&=0,
\end{align}
where we have used the gamma-matrix formulae \eqref{gamma_axial} with $\mathrm{Tr}\qty(\gamma^j\gamma_5P_-\gamma^0)=0$.

We next consider the higher-order terms with $n\geq 3$. 
These terms are not eliminated solely by the Lorentz trace identities. 
We therefore derive a necessary condition for a fixed term in $\xi_{s,r}^{\qty(n)}\qty(M)$ to give a nonvanishing contribution to the fermion-number variation. 
Assuming that a term in the spatial integral contains a factor $M^\beta k_1^{c_1}k_2^{c_2}k_3^{c_3}\ \qty(\beta,c_i\in\mathbb{Z}_{\geq 0})$, the integrals in $\xi_{s,r}^{\qty(n)}\qty(M)$ yield
\begin{align}
	&M^\beta\int\dd[3]{k}\int_0^\infty\dd{t}t^{\frac{s-1}{2}+n}e^{-t\qty(k^2+M^2)}k_1^{c_1}k_2^{c_2}k_3^{c_3}\nonumber\\
	&=\mathcal{G}\qty(c_1,c_2,c_3)\Gamma\qty(\frac{s-2-c}{2}+n)M^{-\qty(s-2-c-\beta+2n)},
\end{align}
where $\mathcal{G}\qty(c_1,c_2,c_3)\equiv\prod_{i=1}^3\frac{1+\qty(-1)^{c_i}}{2}\Gamma\qty(\frac{1+c_i}{2})$ and $c\equiv\sum_{i=1}^3c_i$ (see Appendix~\ref{intformulae}). 
In particular, a nonzero momentum integral requires each $c_i$ to be even; hence, $c$ is a nonnegative even integer. 
After integrating over $M$, a fixed term has the following $s$-dependence:
\begin{align}
	\int_{M_0}^\infty\dd{M}\xi_{s,r}^{\qty(n)}\qty(M)\propto \frac{\Gamma\qty(\frac{s-2-c}{2}+n)}{\Gamma\qty(\frac{s+1}{2})}\frac{sM_0^{-\qty(s-3-c-\beta+2n)}}{s-3-c-\beta+2n}.
\end{align}
Thus, a fixed term can produce a nonzero contribution as $s\to 0^+$ only if 
\begin{align}
	2n-3=c+\beta.\label{xifinite}
\end{align}
This is the necessary condition for $\xi_{s,r}^{\qty(n)}\qty(M)$ to contribute to the fermion-number variation.

We now show that this condition cannot be satisfied. 
To this end, we consider a term in 
\begin{align}
	\qty(K'+V)^{p_0}\prod_{r'=1}^r\qty[K\qty(K'+V)^{p_{r'}}],\ p_0+\sum_{r'=1}^rp_{r'}=n-r.
\end{align}
First, we examine the factor $\qty(K'+V)^{p_0}$. 
Since this factor contains $p_0$ powers of $\qty(K'+V)$, the degree in $k_j$ and $M$ of any term in its expansion cannot exceed $p_0$. 
Therefore, if such a term contains $M^{\beta_0}k_1^{c_1^0}k_2^{c_2^0}k_3^{c_3^0}\ \qty(\beta_0,c_i^0\in\mathbb{Z}_{\geq 0})$ with $c^0\equiv\sum_{i=1}^3c_i^0$, the degree $c^0+\beta_0$ satisfies 
\begin{align}
	0\leq c^0+\beta_0\leq p_0.
\end{align}
The same argument applies to each factor $\qty(K'+V)^{p_{r'}}$.  
Specifically, if a term in $\qty(K'+V)^{p_{r'}}$ contains $M^{\beta_{r'}}k_1^{c_1^{r'}}k_2^{c_2^{r'}}k_3^{c_3^{r'}}\ \qty(\beta_{r'},c_i^{r'}\in\mathbb{Z}_{\geq 0})$ with $c^{r'}\equiv\sum_{i=1}^3c_i^{r'}$, the degree $c^{r'}+\beta_{r'}$ satisfies 
\begin{align}
	0\leq c^{r'}+\beta_{r'}\leq p_{r'}.
\end{align}
Meanwhile, the explicit factors of $K$ carry neither $k_j$ nor $M$, and hence they do not affect the total degree. 
Consequently, for any term in the whole product, the total degree 
\begin{align}
	c+\beta=c^0+\beta_0+\sum_{r'=1}^r\qty(c^{r'}+\beta_{r'})
\end{align}
satisfies
\begin{align}
	0\leq c+\beta\leq n-r.
\end{align}
Since $r\geq 1$, this implies $c+\beta\leq n-1$. 
For $n\geq 3$, however, the necessary condition~\eqref{xifinite} requires $c+\beta=2n-3$, which is incompatible with $c+\beta\leq n-1$. 
Therefore, no term in $\xi_{s,r}^{\qty(n)}\qty(M)$ for $n\geq 3$ can satisfy the necessary condition~\eqref{xifinite}. 
Accordingly, within the term-by-term treatment, we conclude that
\begin{align}
	\lim_{s\to 0^+}\frac{1}{2}\int_{M_0}^\infty\dd{M}\xi_s\qty(M)=0.
\end{align}

Next, we consider the contributions from $\delta_s\qty(M)$. 
For the two lowest orders $n=0,1$, we find
\begin{subequations}
\begin{align}
	\delta_{s,l,\vb{m}}^{\qty(0)}\qty(M)
	&=\frac{\qty(-1)s}{\qty(2\pi)^3\Gamma\qty(\frac{s+1}{2})}\int\dd[3]{k}\int_0^\infty\dd{t}t^{\frac{s-1}{2}}e^{-t\qty(k^2+M^2)}\int\dd[3]{x}\mathrm{Tr}\qty[\frac{\qty(-t)^l}{l!}\Delta^l\gamma^0]\nonumber\\
	&=0,
\end{align}
\begin{align}
	\delta_{s,l,\vb{m}}^{\qty(1)}\qty(M)
	&=\frac{s}{\qty(2\pi)^3\Gamma\qty(\frac{s+1}{2})}\int\dd[3]{k}\int_0^\infty\dd{t}t^{\frac{s-1}{2}}e^{-t\qty(k^2+M^2)}\nonumber\\
	&\times\int\dd[3]{x}\mathrm{Tr}\qty[F_{l,\vb{m}}^{\qty(1)}\qty(t)\Delta^l\mathrm{ad}_\Delta^{m_1}\qty(K'+V+K)\gamma^0]\nonumber\\
	&=0,
\end{align}
\end{subequations}
Here, we have used the Lorentz trace identities~\eqref{gamma_axial} and also the fact that $\Delta$ carries no gamma matrices; thus, the commutator $\mathrm{ad}_\Delta^{m_1}$ does not change the gamma-matrix structure of $K'+V+K$.

For $n\geq 2$, however, the Lorentz trace identities alone do not eliminate the terms. 
We therefore derive a necessary condition for a fixed term in $\delta_{s,l,\vb{m}}^{\qty(n)}\qty(M)$ to give a finite contribution to the fermion-number variation. 
Let $m\equiv \sum_{b=1}^nm_b$. 
Since $F_{l,\vb{m}}^{\qty(n)}\qty(t)$ is proportional to $t^{l+m+n}$, a term containing the factor $M^\beta k_1^{c_1}k_2^{c_2}k_3^{c_3}\ \qty(\beta,c_i\in\mathbb{Z}_{\geq 0})$ gives 
\begin{align}
	&M^\beta\int\dd[3]{k}\int_0^\infty\dd{t}t^{\frac{s-1}{2}}e^{-t\qty(k^2+M^2)}t^{l+m+n}k_1^{c_1}k_2^{c_2}k_3^{c_3}\nonumber\\
	&=\mathcal{G}\qty(c_1,c_2,c_3)\Gamma\qty(\frac{s-2-c+2l+2m}{2}+n)M^{-\qty(s-2-c+2l+2m+2n-\beta)}.
\end{align}
Here $\mathcal{G}\qty(c_1,c_2,c_3)\equiv\prod_{i=1}^3\frac{1+\qty(-1)^{c_i}}{2}\Gamma\qty(\frac{1+c_i}{2})$ and $c\equiv\sum_{i=1}^3c_i$ (see Appendix~\ref{intformulae}). 
A nonzero momentum integral requires each $c_i$ to be even; hence, $c$ is a nonnegative even integer. 
After the $M$-integration, the $s$-dependence of such a term becomes
\begin{align}
        \int_{M_0}^\infty\dd{M}\delta_{s,l,\vb{m}}^{\qty(n)}\qty(M)\propto \frac{\Gamma\qty(\frac{s-2-c+2l+2m}{2}+n)}{\Gamma\qty(\frac{s+1}{2})}\frac{sM_0^{-\qty(s-3-c+2l+2m+2n-\beta)}}{s-3-c+2l+2m+2n-\beta}.
\end{align}
Therefore, a contribution can remain nonzero as $s\to 0^+$ only if 
\begin{align}
	2l+2m+2n-3=c+\beta.\label{deltafinite2}
\end{align}
This condition is necessary for $\delta_{s,l,\vb{m}}^{\qty(n)}\qty(M)$ to contribute to $\Delta N_{\mathrm{F}}$.

We next estimate the possible values of $c+\beta$. 
As in the previous section, the factor $\Delta^l$ can generate at most $l$ powers of momentum $k_j$. 
Similarly, for each factor $\mathrm{ad}_\Delta^{m_b}\qty(K'+V+K)$, $m_b$ commutators with $\Delta$ can generate at most $m_b$ additional powers of $k_j$. 
The operator on which these commutators act also depends on $k_j$ and $M$: $K'$ carries one power of $k_j$, the mass-dependent part of $V$ carries one power of $M$, while $K$ and the remaining part of $V$ carry neither $k_j$ nor $M$. 
Hence each factor $\mathrm{ad}_\Delta^{m_b}\qty(K'+V+K)$ can contribute at most $m_b+1$ powers to the total degree $c+\beta$. 
Therefore, any term in $\Delta^l\prod_{b=1}^n\mathrm{ad}_\Delta^{m_b}\qty(K'+V+K)$ satisfies 
\begin{align}
	c+\beta\leq l+\sum_{b=1}^n\qty(m_b+1)=l+m+n.
\end{align}
Combining this bound with the necessary condition~\eqref{deltafinite2} yields
\begin{align}
	2l+2m+2n-3\leq l+m+n
\end{align}
or equivalently
\begin{align}
	l+m+n\leq 3.
\end{align}
Since $l+m>0$ by definition of $\delta_s\qty(M)$, and since we are considering only $n\geq 2$, the only possible case is 
\begin{align}
	n=2,\ l+m=1.
\end{align}
Thus, all terms with $n\geq 3$ are excluded based on the degree-counting argument above.

It remains to examine the exceptional case $n=2$ with $l+m=1$. 
This case has two possibilities: $\qty(l,m)=\qty(1,0),\qty(0,1)$. 
The corresponding terms are of the form
\begin{subequations}
\begin{align}
	&\delta_{s,1,\qty(0,0)}^{\qty(2)}\qty(M)
        =\frac{\qty(-1)^{n+1}s}{\qty(2\pi)^3\Gamma\qty(\frac{s+1}{2})}\int\dd[3]{k}\int_0^\infty\dd{t}t^{\frac{s-1}{2}}e^{-t\qty(k^2+M^2)}\nonumber\\
	&\times\int\dd[3]{x}\mathrm{Tr}\qty[F_{1,\qty(0,0)}^{\qty(2)}\qty(t)\Delta\qty(K'+V+K)^2\gamma^0]
\end{align}
\begin{align}
	\delta_{s,0,\qty(1,0)}^{\qty(2)}\qty(M)+\delta_{s,0,\qty(0,1)}^{\qty(2)}\qty(M)
        &=\frac{\qty(-1)^{n+1}s}{\qty(2\pi)^3\Gamma\qty(\frac{s+1}{2})}\int\dd[3]{k}\int_0^\infty\dd{t}t^{\frac{s-1}{2}}e^{-t\qty(k^2+M^2)}\nonumber\\
	&\times\int\dd[3]{x}\mathrm{Tr}\qty[F_{0,\qty(1,0)}^{\qty(2)}\qty(t)\mathrm{ad}_\Delta\qty(K'+V+K)\qty(K'+V+K)\gamma^0]\nonumber\\
        &+\frac{\qty(-1)^{n+1}s}{\qty(2\pi)^3\Gamma\qty(\frac{s+1}{2})}\int\dd[3]{k}\int_0^\infty\dd{t}t^{\frac{s-1}{2}}e^{-t\qty(k^2+M^2)}\nonumber\\
	&\times\int\dd[3]{x}\mathrm{Tr}\qty[F_{0,\qty(0,1)}^{\qty(2)}\qty(t)\qty(K'+V+K)\mathrm{ad}_\Delta\qty(K'+V+K)\gamma^0].
\end{align}
\end{subequations}
Only terms that saturate the bound $c+\beta\leq l+m+n=3$ can satisfy the necessary condition~\eqref{deltafinite2}. 
Since a nonzero momentum integral requires $c$ to be a nonnegative even integer, the possible pairs are 
\begin{align}
	\qty(c,\beta)=\qty(0,3),\ \qty(2,1).
\end{align}
The first possibility, $\qty(c,\beta)=\qty(0,3)$, cannot occur in the present case because there are only two factors of $K'+V+K$, and hence at most two powers of $M$ can be produced. 
Therefore, the only remaining possibility is $\qty(c,\beta)=\qty(2,1)$. 
Let us identify the terms that can realize this possibility. 
Since $l+m=1$, the factor $\Delta$ or the commutator $\mathrm{ad}_\Delta$ can supply at most one power of $k_j$. 
Thus, in order to obtain $c=2$, one additional power of $k_j$ must arise from the two factors of $K'+V+K$. 
Meanwhile, the condition $\beta=1$ requires one of these two factors to be the mass-dependent component of $V$, namely $M\gamma^j\gamma_5A_j$. 
Thus, the remaining factor must be $K'$; otherwise the total momentum degree would be at most one, and the condition $c=2$ could not be satisfied. 
Hence the only relevant structures are the cross-terms between $K'$ and $M\gamma^j\gamma_5A_j$. 
The extra momentum, supplied by $\Delta$ or $\mathrm{ad}_\Delta$, carries no gamma matrices. 
Then, the Lorentz traces of the relevant terms reduce to 
\begin{align}
	\mathrm{Tr}\qty(P_-\gamma^j\gamma_5\gamma^0)=0,\ \mathrm{Tr}\qty(\gamma^j\gamma_5P_-\gamma^0)=0.
\end{align}
Therefore, the exceptional terms with $n=2$ and $l+m=1$ do not contribute to the fermion-number variation, and so we conclude that
\begin{align}
	\lim_{s\to 0^+}\frac{1}{2}\int_{M_0}^\infty\dd{M}\delta_s\qty(M)=0.
\end{align}

Together with the result for $\xi_s\qty(M)$,
\begin{align*}
	\lim_{s\to 0^+}\frac{1}{2}\int_{M_0}^\infty\dd{M}\xi_s\qty(M)=0,
\end{align*}
the remaining terms give no contribution within the term-by-term analysis. 
Combining this result with the finite contribution obtained in the previous subsection, we finally find
\begin{align}
	\Delta N_{\mathrm{F}}=Q
\end{align}
within the term-by-term treatment. 
Thus, the axial-vector coupling obtained by the $\mathrm{SU}\qty(N)$ chiral rotation also yields the same fermion-number variation as the standard chiral Yukawa coupling. 
In particular, the additional operator $K$ and the remaining corrections in the expansion do not modify the final result.

\subsection{Vector coupling}
\label{Comp22}
We next consider the following Lagrangian with a vector coupling
\begin{align}
	\mathcal{L}=\overline{\Psi}\qty(i\gamma^\mu D_\mu-M)\Psi,\ D_\mu=\partial_\mu-iA_\mu\label{vec}
\end{align}
where $A_\mu=iU^\dagger\partial_\mu U$. 
For a static background with $A_0=0$, the corresponding Dirac Hamiltonian is 
\begin{align}
	\mathcal{H}=-\gamma^0\qty(i\gamma^jD_j-M).\label{Hvec}
\end{align}
This model can be obtained from the axial-vector coupling~\eqref{CQSMrot} by replacing the chiral projection operator $P_-$ with the identity matrix. 
This model provides a useful comparison with the axial-vector coupling. 

Before presenting the explicit calculation, we summarize the roadmap. 
The crucial difference from the axial-vector coupling is that the squared Hamiltonian contains no mass-dependent interaction with the background gauge field. 
Thus, in the degree-counting analysis, the power $\beta$ generated by the interaction terms is always zero. 
Consequently, the necessary conditions for a fixed term to contribute to the fermion-number variation require an odd total momentum degree, whereas a nonvanishing momentum integral requires an even total momentum degree. 
We therefore expect each contribution to vanish term by term. 
We first confirm this for the lowest orders and then apply the degree-counting argument to general order. 

To compute the fermion-number variation $\Delta N_{\mathrm{F}}$, we first evaluate $\mathcal{H}^2$ and $\pdv{\mathcal{H}}{M}$. 
They are expressed as 
\begin{subequations}
\begin{align}
	\mathcal{H}^2
	&=\mathcal{H}_0^2+K+V,\ \pdv{\mathcal{H}}{M}=\gamma^0,
\end{align}
\begin{align}
	\mathcal{H}_0^2\equiv -\partial_j^2+M^2,\ 
	K\equiv 2iA_j\partial_j,\ 
	V\equiv -i\gamma^j\gamma^kD_jA_k.
\end{align}
\end{subequations}
Note that the interaction term $V$ does not contain the mass-parameter $M$. 
Following the same procedure as in the previous subsections, we obtain the formal decomposition
\begin{align}
	\pdv{\eta\qty(s;\mathcal{H})}{M}=\mathcal{Q}_s\qty(M)+\xi_s\qty(M)+\delta_s\qty(M).\label{etanvec}
\end{align}
Here we define
\begin{subequations}
\begin{align}
	\mathcal{Q}_s\qty(M)
	&=\sum_{n=0}^\infty\mathcal{Q}_s^{\qty(n)}\qty(M),
\end{align}
\begin{align}
	\mathcal{Q}_s^{\qty(n)}\qty(M)
	&\equiv \frac{\qty(-1)^{n+1}s}{n!\qty(2\pi)^3\Gamma\qty(\frac{s+1}{2})}\int\dd[3]{k}\int_0^\infty\dd{t}t^{\frac{s-1}{2}+n}e^{-t\qty(k^2+M^2)}\nonumber\\
	&\times\int\dd[3]{x}\mathrm{Tr}\qty[\qty(-2A_jk_j-i\gamma^j\gamma^kD_jA_k)^n\gamma^0],
\end{align}
\begin{align}
	\xi_s\qty(M)
	&=\sum_{n=0}^\infty\sum_{r=1}^n\xi_{s,r}^{\qty(n)}\qty(M),\nonumber
\end{align}
\begin{align}
	&\xi_{s,r}^{\qty(n)}\qty(M)\equiv\frac{\qty(-1)^{n+1}s}{n!\qty(2\pi)^3\Gamma\qty(\frac{s+1}{2})}\int\dd[3]{k}\int_0^\infty\dd{t}t^{\frac{s-1}{2}+n}e^{-t\qty(k^2+M^2)}\nonumber\\
	&\times\int\dd[3]{x}\mathrm{Tr}\qty[\sum_{p_0+\cdots+p_r=n-r}\qty(K'+V)^{p_0}\prod_{r'=1}^r\qty[K\qty(K'+V)^{p_{r'}}]\gamma^0],
\end{align}
\begin{align}
	\delta_s\qty(M)=\sum_{n=0}^\infty\sum_{l+\sum_{b=1}^nm_b>0}\delta_{s,l,\vb{m}}^{\qty(n)}\qty(M),\nonumber
\end{align}
\begin{align}
	&\delta_{s,l,\vb{m}}^{\qty(n)}\qty(M)\equiv\frac{\qty(-1)^{n+1}s}{\qty(2\pi)^3\Gamma\qty(\frac{s+1}{2})}\int\dd[3]{k}\int_0^\infty\dd{t}t^{\frac{s-1}{2}}e^{-t\qty(k^2+M^2)}\nonumber\\
	&\times\int\dd[3]{x}\mathrm{Tr}\qty[F_{l,\vb{m}}^{\qty(n)}\qty(t)\Delta^l\prod_{b=1}^n\mathrm{ad}_\Delta^{m_b}\qty(K'+V+K)\gamma^0],
\end{align}
\end{subequations}
where $K'\equiv -2A_jk_j$. 

We first consider the lowest orders. 
For $n=0,1$, the relevant terms vanish because of the Lorentz trace identities. 
In particular, we have
\begin{subequations}
\begin{align}
	\mathcal{Q}_s^{\qty(0)}\qty(M)
	&=\frac{\qty(-1)s}{\qty(2\pi)^3\Gamma\qty(\frac{s+1}{2})}\int\dd[3]{k}\int_0^\infty\dd{t}t^{\frac{s-1}{2}}e^{-t\qty(k^2+M^2)}\int\dd[3]{x}\mathrm{Tr}\qty(\gamma^0)\nonumber\\
	&=0,
\end{align}
\begin{align}
	\mathcal{Q}_s^{\qty(1)}\qty(M)
	&=\frac{s}{\qty(2\pi)^3\Gamma\qty(\frac{s+1}{2})}\int\dd[3]{k}\int_0^\infty\dd{t}t^{\frac{s-1}{2}+1}e^{-t\qty(k^2+M^2)}\nonumber\\
	&\times\int\dd[3]{x}\mathrm{Tr}\qty[\qty(-2A_jk_j-i\gamma^j\gamma^kD_jA_k)\gamma^0]\nonumber\\
	&=0,
\end{align}
\begin{align}
	\xi_{s,1}^{\qty(1)}\qty(M)
	&\equiv\frac{s}{\qty(2\pi)^3\Gamma\qty(\frac{s+1}{2})}\int\dd[3]{k}\int_0^\infty\dd{t}t^{\frac{s-1}{2}+1}e^{-t\qty(k^2+M^2)}
	\int\dd[3]{x}\mathrm{Tr}\qty[K\gamma^0]\nonumber\\
	&=0,
\end{align}
\begin{align}
	\delta_{s,l,\vb{m}}^{\qty(0)}\qty(M)
	&\equiv\frac{\qty(-1)s}{\qty(2\pi)^3\Gamma\qty(\frac{s+1}{2})}\int\dd[3]{k}\int_0^\infty\dd{t}t^{\frac{s-1}{2}}e^{-t\qty(k^2+M^2)}
	\int\dd[3]{x}\mathrm{Tr}\qty[\frac{\qty(-t)^l}{l!}\Delta^l\gamma^0]\nonumber\\
	&=0,
\end{align}
\begin{align}
	\delta_{s,l,\vb{m}}^{\qty(1)}\qty(M)
	&\equiv\frac{s}{\qty(2\pi)^3\Gamma\qty(\frac{s+1}{2})}\int\dd[3]{k}\int_0^\infty\dd{t}t^{\frac{s-1}{2}}e^{-t\qty(k^2+M^2)}\nonumber\\
	&\times\int\dd[3]{x}\mathrm{Tr}\qty[F_{l,\vb{m}}^{\qty(1)}\qty(t)\Delta^l\mathrm{ad}_\Delta^{m_1}\qty(K'+V+K)\gamma^0]\nonumber\\
	&=0.
\end{align}
\end{subequations}
Here we have used $\mathrm{Tr}\qty(\gamma^0)=0,\ \mathrm{Tr}\qty(\gamma^j\gamma^k\gamma^0)=0$. 
Note that the $n=0$ contribution to $\xi_s\qty(M)$ is absent because the sum over $r$ is empty.

For $n>1$, however, these trace identities alone do not necessarily eliminate the terms at the integrand level. 
These terms nevertheless vanish after the $M$-integration and the subsequent $s\to 0^+$ limit. 
Since $\beta=0$, the necessary conditions obtained from the $k$-and $t$-integrals reduce to
\begin{subequations}
\begin{align}
	\mathcal{Q}_s^{\qty(n)}\ &:\ 2n-3=a\label{finite21}\\
	\xi_{s,r}^{\qty(n)}\ &:\ 2n-3=c,\label{finite22}\\
	\delta_{s,l,\vb{m}}^{\qty(n)}\ &:\ 2l+2m+2n-3=c.\label{finite23}
\end{align}
\end{subequations}
Here $a$ and $c$ denote the total powers of momentum in the corresponding momentum integrals. 
A nonzero momentum integral requires these powers to be nonnegative even integers. 
By contrast, the left-hand sides of Eqs.~\eqref{finite21}-\eqref{finite23} are odd integers. 
Hence, none of the necessary conditions can be satisfied. 
Therefore, every fixed term with $n>1$ vanishes after the $M$-integration and the subsequent $s\to0^+$ limit.

Combining the lowest order analysis with the above necessary condition argument, we obtain, within the term-by-term treatment,
\begin{align}
	\Delta N_{\mathrm{F}}=0.
\end{align}
Thus, the vector coupling yields a vanishing fermion-number variation.

For the present pure-gauge configuration $A_\mu=iU^\dagger\partial_\mu U$, this result has a simple interpretation. 
The vector-coupled Hamiltonian $\mathcal{H}$~\eqref{Hvec} is unitarily equivalent to the massive-free Dirac Hamiltonian $\mathcal{H}_0$, 
\begin{align}
	\mathcal{H}=U^\dagger\mathcal{H}_0U,\ \mathcal{H}_0=-\gamma^0\qty(i\gamma^j\partial_j-M).
\end{align}
It follows that $\mathcal{H}$ and $\mathcal{H}_0$ have identical spectra. 
Consequently, in this pure-gauge case, neither background-induced spectral asymmetry nor zero-energy level crossing occurs.

The contrast with the axial-vector coupling is also evident in the squared Hamiltonian. 
In the axial-vector case, the interaction term $V$ in the squared Hamiltonian $\mathcal{H}^2=\mathcal{H}_0^2+K+V$ contains the $M$-dependent term $M\gamma^j\gamma^5A_j$; meanwhile, no analogous $M$-dependent interaction appears in the vector coupling case. 
This structural difference distinguishes the degree-counting arguments in the two cases: the axial-vector coupling admits a term with the required powers of $M$ and momentum, whereas the vector coupling does not. 
Thus, within the coupling schemes considered here, the same topologically nontrivial field $U$ can lead to different fermion-number variations depending on how it is coupled to the fermion.

\subsection{Remarks on general gauge backgrounds}
\label{Remark}

In the preceding analyses, the background gauge field has been restricted to the pure-gauge form $A_j=iU^\dagger\partial_jU$. 
Here, we briefly comment on the case of general static gauge backgrounds in the temporal gauge $A_0=0$.

For the axial-vector coupling case, discussed in Sec.~\ref{Comp21}, the difference from the pure-gauge analysis arises only in the $n=2$ contribution to the finite part $\mathcal{Q}_s^{\qty(n)}\qty(M)$. 
Indeed, the higher-derivative terms in $\mathcal{Q}_s^{\qty(n)}\qty(M)$ with $n>3$, together with $\delta_s\qty(M)$ and $\xi_s\qty(M)$, vanish for an arbitrary gauge field $A_j$ in the same term-by-term treatment. 
In other words, the degree-counting arguments for the higher-order terms do not rely on the pure-gauge condition. 
For a general gauge background, however, the $n=2$ contribution need not vanish and the $n=3$ contribution is also retained. 
These two contributions combine to yield 
\begin{align}
    \Delta N_{\mathrm{F}}=N_{\mathrm{CS}},
\end{align}
where
\begin{align}
    N_{\mathrm{CS}}=-\frac{1}{8\pi^2}\int \dd[3]{x}\epsilon^{jkl}\mathrm{Tr}\qty(A_j\partial_kA_l-\frac{2i}{3}A_jA_kA_l)
\end{align}
is the Chern--Simons functional for the hermitian gauge field $A_j$~\cite{Nakahara:2003nw}. 
For the pure-gauge chiral soliton background, the Chern--Simons functional reduces to the winding number of $U$,
\begin{align}
    N_{\mathrm{CS}}\vert_{A_j=iU^\dagger\partial_jU}=Q,
\end{align}
thereby recovering the result obtained in Sec.~\ref{Comp21}. 
This result shows that the relation $\Delta N_{\mathrm{F}}=Q$ for the pure-gauge chiral soliton background naturally extends to $\Delta N_\mathrm{F}=N_{\mathrm{CS}}$ for general gauge backgrounds. 
For a generic gauge background, the Chern--Simons functional is not necessarily integer-valued. 
Thus, the integer-valued fermion-number variation $\Delta N_{\mathrm{F}}=Q$ in the chiral-soliton case follows from the pure-gauge form of the background, for which the Chern--Simons functional then reduces to the winding number.

Finally, we emphasize that the result within the term-by-term degree-counting argument in the vector coupling case, discussed in Sec.~\ref{Comp22}, does not depend on the pure-gauge condition $A_j=iU^\dagger\partial_jU$, i.e., the result $\Delta N_{\mathrm{F}}=0$ remains valid for arbitrary static gauge backgrounds. 
This result is also consistent with the spectral symmetry of the vector-coupled Hamiltonian: its positive- and negative-energy spectra are symmetric, so that the regularized spectral asymmetry vanishes. 
Unlike the pure-gauge case, however, a general gauge background can modify the fermionic spectrum while preserving this symmetry.

\section{Summary and Discussion}
\label{Summary}

In this work, we have investigated whether the fermion-number response to a topologically nontrivial chiral soliton background is determined solely by the topology or also by the fermion--soliton coupling scheme. 
Starting from the chiral quark soliton model (CQSM)~\eqref{CQSM}, we formulate the regularized fermion number in terms of the $\eta$-invariant and evaluate its variation along the dynamical quark mass parameter space. 
The resulting fermion-number variation reproduces the topological charge:
\begin{align}
	\Delta N_{\mathrm{F}}=Q,
\end{align}
where $Q$ is the topological charge of the chiral field $U\in \mathrm{SU}\qty(N)$. 
The topological contribution arises from the third-order term in the expansion, whereas the remaining terms vanish term by term after the $M$-integration and the subsequent $s\to0^+$ limit. 
This result is consistent with the well-known level-crossing picture of the CQSM, in which the net number of level crossings reflects the topological charge of the chiral soliton. 

We further examine two alternative fermion--background couplings to clarify which coupling features are responsible for producing the fermion-number variation. 
The axial-vector coupling, obtained from the standard CQSM by a chiral rotation, likewise yields
\begin{align}
	\Delta N_{\mathrm{F}}=Q.
\end{align}
By contrast, the vector coupling obtained by replacing the chiral projection operator with the identity matrix in the axial-vector coupling yields a vanishing fermion-number variation 
\begin{align}
	\Delta N_{\mathrm{F}}=0.
\end{align}
The results for the three coupling schemes analyzed in this work are summarized in Table~\ref{tab:sum}.
\begin{table}[tbp]
        \centering
                \begin{tabular}{|c|cc|}
                        \hline
                        Coupling
			&Hamiltonian
			&Fermion-number variation\\
                        \hline 
                        $\overline{\Psi}U^{\gamma_5}\Psi$
			&$\mathcal{H}_{\mathrm{CQSM}}\equiv-\gamma^0\qty(i\gamma^j\partial_j-MU^{\gamma_5})$
			&$\Delta N_{\mathrm{F}}=Q$\\
			$\overline{\Psi}\gamma^\mu P_-A_\mu\Psi$
			&$\mathcal{H}_{\mathrm{axial}}\equiv-\gamma^0\qty[i\gamma^j\qty(\partial_j-iP_-A_j)-M]$
			&$\Delta N_{\mathrm{F}}=Q$\\
			$\overline{\Psi}\gamma^\mu A_\mu\Psi$
			&$\mathcal{H}_{\mathrm{vector}}\equiv-\gamma^0\qty[i\gamma^j\qty(\partial_j-iA_j)-M]$
			&$\Delta N_{\mathrm{F}}=0$\\
                        \hline
                \end{tabular}
        \caption{\label{tab:sum} Comparison of the fermion-number variation for the three fermion--chiral soliton coupling schemes considered in this work. 
	The chiral Yukawa coupling of the CQSM and the axial-vector coupling exhibit a nontrivial fermion number variation equal to the topological charge $Q$, whereas the vector coupling for the pure-gauge configuration $A_j=iU^\dagger\partial_jU$ yields a vanishing fermion-number variation. 
	}
\end{table}
The comparison demonstrates that the nontrivial winding of the chiral field alone is insufficient to determine the fermion-number variation. 
The same chiral field $U$, carrying the same topological charge $Q$, produces different fermionic responses depending on how it is coupled to the fermion. 
Thus, the topology--fermion-number correspondence depends on the coupling scheme as well as on the topology of the background itself. 

The vanishing fermion-number variation for the vector coupling can be understood as a consequence of unitary equivalence to the massive-free Hamiltonian. 
Let $\mathcal{H}_0$ denote the massive-free Dirac Hamiltonian. 
If a Hamiltonian $\mathcal{H}$ can be expressed as $\mathcal{H}=W^\dagger \mathcal{H}_0W$ using a unitary transformation $W$, then $\mathcal{H}$ has the same spectrum as $\mathcal{H}_0$ for each $M$. 
Since the massive-free Hamiltonian does not couple to the chiral soliton background, such a unitary equivalence produces no background-induced spectral asymmetry in the spectrum of $\mathcal{H}$. 
For a pure-gauge configuration $A_j=iU^\dagger\partial_jU$, the vector-coupled Hamiltonian satisfies 
\begin{align}
	\mathcal{H}_{\mathrm{vector}}=U^\dagger \mathcal{H}_0U.\label{unitary}
\end{align}
Thus, even when the chiral field $U$ carries a nontrivial topological charge, the vector coupling represents merely a local flavor rotation of the massive-free fermion and cannot induce nontrivial fermion number variation.

As discussed in Sec.~\ref{Remark}, the vanishing of the fermion-number variation in the vector coupling is not restricted to the pure-gauge case and also holds for a general static vector gauge field within our formulation. 
Although such backgrounds can modify the fermionic spectrum, the symmetry between positive- and negative-energy states ensures that the regularized spectral asymmetry vanishes. 
Nevertheless, only in the pure-gauge case is the Hamiltonian unitarily equivalent to the massive-free Dirac Hamiltonian, as shown in Eq.~\eqref{unitary}.

For the axial-vector coupling with a general static gauge background in temporal gauge, the same analysis yields 
\begin{align}
    \Delta N_{\mathrm{F}} = N_{\mathrm{CS}},
\end{align}
where $N_{\mathrm{CS}}$ denotes the Chern--Simons functional. 
For the pure-gauge configuration $A_j=iU^\dagger\partial_jU$, this functional reduces to the topological charge $N_{\mathrm{CS}}=Q$. 

We emphasize that the analytical results in Sec.~\ref{Comp} and Sec.~\ref{Comp2} are based on a term-by-term treatment of the expansion. 
More precisely, each term in the expansion of the non-topological sectors vanishes after the $M$-integration and the subsequent $s\to0^+$ limit. 
We do not establish convergence or resummation of the full infinite series. 
The three coupling schemes have been analyzed within the same framework, allowing their fermion-number responses to be compared on a common basis. 

The present study clarifies an important point: the topology of the background field alone is insufficient to produce a nontrivial fermion number. 
What matters is how the background field enters the fermionic Hamiltonian. 
In the vector-coupling case, the chiral field $U$ enters only through a pure-gauge vector connection, which can be eliminated by a unitary transformation. 
Therefore, the fermionic spectrum is identical to that of the massive-free Hamiltonian. 
By contrast, neither the CQSM Yukawa coupling $\mathcal{H}_{\mathrm{CQSM}}$ nor the axial-vector coupling $\mathcal{H}_{\mathrm{axial}}$ can be reduced to the massive-free Hamiltonian by a vector-like flavor rotation.

Finally, the present analysis suggests several directions for future investigation. 
A particularly important problem is to identify general criteria for determining whether a given coupling scheme produces a nontrivial fermion number. 
Such a classification would systematically clarify how soliton topology is encoded in fermionic spectra and may provide a framework for investigating novel fermion--soliton systems. 
A further direction is to extend this approach to more general coupling schemes involving all possible spinor bilinears. 

\appendix
\section{Integral formula}
\label{intformulae}

In this appendix, we derive an integral formula 
\begin{align}
	&\int\dd[3]{k}\int_0^\infty\dd{t} t^{\frac{s-1}{2}+n}e^{-t\qty(k^2+M^2)}k_1^{a_1}k_2^{a_2}k_3^{a_3}
	=\mathcal{G}\qty(a_1,a_2,a_3)\Gamma\qty(\frac{s-2-a}{2}+n)M^{-\qty(s-2-a+2n)}\label{inta}
\end{align}
with $\mathcal{G}\qty(a_1,a_2,a_3)\equiv\prod_{i=1}^3\frac{1+\qty(-1)^{a_i}}{2}\Gamma\qty(\frac{1+a_i}{2})$ and $a\equiv\sum_{i=1}^3a_i$. 
This integral is directly used in Sec.~\ref{Comp2}. 
If one takes $a=a_1=a_2=a_3=0$, Eq.~\eqref{inta} becomes 
\begin{align}
	&\int\dd[3]{k}\int_0^\infty\dd{t} t^{\frac{s-1}{2}+n}e^{-t\qty(k^2+M^2)}
	=\pi^{\frac{3}{2}}\Gamma\qty(\frac{s-2}{2}+n)M^{-\qty(s-2+2n)}.\label{int}
\end{align}
This integral is used in Sec.~\ref{Comp}. 
Since Eq.~\eqref{int} is obtained from Eq.~\eqref{inta}, we derive Eq.~\eqref{inta} only.

The strategy is to follow the steps below:
\begin{enumerate}
	\item Separate variables $t$ and $k$ in the integral,
	\item Perform the Gaussian integral with respect to $k$,
	\item Use the integral representation of the gamma function:
	
	$\Gamma\qty(p)=\int_0^\infty\dd{t} t^{p-1}e^{-t}$.
\end{enumerate}
The details are as follows:
\begin{align}
	\int\dd[3]{k}\int_0^\infty\dd{t} t^{\frac{s-1}{2}+n}e^{-t\qty(k^2+M^2)}k_1^{a_1}k_2^{a_2}k_3^{a_3}
	&=\int_0^\infty\dd{t} t^{\frac{s-1}{2}+n}e^{-tM^2}\int\dd[3]{k}e^{-tk^2}k_1^{a_1}k_2^{a_2}k_3^{a_3}\nonumber\\
	&=\int_0^\infty\dd{t} t^{\frac{s-4-a}{2}+n}e^{-tM^2}\prod_{i=1}^3\int\dd{k_i}e^{-k_i^2}k_i^{a_i}\nonumber\\
	&=\mathcal{G}\qty(a_1,a_2,a_3)\Gamma\qty(\frac{s-2-a}{2}+n)M^{-\qty(s-2-a+2n)}.
\end{align}
The generalization for $d$-dimensional space is straightforward.

\acknowledgments
This work is supported in part by JSPS KAKENHI [Grants No. JP23K02794 (NS)] and JST SPRING [Grant Number JPMJSP2151 (SY)]. 

\bibliographystyle{JHEP}
\bibliography{ref_sf_CQSM}

\end{document}